\documentclass[fleqn,usenatbib]{mnras}

\usepackage{newtxtext,newtxmath}
\usepackage{comment}
\usepackage{listings}
\usepackage[T1]{fontenc}
\usepackage{xcolor, color}

\DeclareRobustCommand{\VAN}[3]{#2}
\let\VANthebibliography\thebibliography
\def\thebibliography{\DeclareRobustCommand{\VAN}[3]{##3}\VANthebibliography}

\usepackage{graphicx}	
\usepackage{amsmath}	
\usepackage{bm}
\usepackage{physics}
\usepackage{tabularx}
\usepackage[inline]{enumitem}
\usepackage{parskip}
\newcommand{\tempoDOS}{$\mathrm{{\scriptstyle TEMPO2}}$}

\title[Population-level estimation of torque-law braking indices]{Hierarchical Bayesian estimation of population-level torque law parameters from anomalous pulsar braking indices}

\author[A.~F.~Vargas et al.]{\parbox{\linewidth}{\centering
Andr\'es F. Vargas$^{1,2}$\thanks{E-mail: a.vargas@unimelb.edu.au}, Julian B. Carlin$^{1,2,3}$, and
Andrew Melatos$^{1,2}$
}\\\\
$^{1}$School of Physics, University of Melbourne, Parkville, VIC 3010, Australia\\
$^{2}$OzGrav: The Australian Research Council Centre of Excellence for Gravitational-wave Discovery, University of Melbourne, Parkville, VIC 3010, Australia\\
$^{3}$Department of Infectious Diseases, University of Melbourne, Parkville, VIC 3010, Australia}

\date{Accepted XXX. Received YYY; in original form ZZZ}

\pubyear{2024}

\begin{document}
\label{firstpage}
\pagerange{\pageref{firstpage}--\pageref{lastpage}}
\maketitle

\begin{abstract}

\noindent Stochastic fluctuations in the spin frequency $\nu$ of a rotation-powered pulsar affect how accurately one measures the power-law braking index, $n_{\rm pl}$, defined through $\dot{\nu}=K\nu^{n_{\rm pl}}$, and can lead to measurements of anomalous braking indices, with $\vert n \vert = \vert \nu \ddot{\nu}/ \dot{\nu}^{2} \vert \gg1$, where the overdot symbolizes a derivative with respect to time. Previous studies show that the variance of the measured $n$ obeys the predictive, falsifiable formula $\langle n^{2} \rangle = n_{\rm pl}^{2}+\sigma^{2}_{\ddot{\nu}}\nu^{2}\gamma_{\ddot{\nu}}^{-2}\dot{\nu}^{-4}T_{\rm obs}^{-1}$ for $\dot{K}=0$, where $\sigma_{\ddot{\nu}}$ is the timing noise amplitude, $\gamma_{\ddot{\nu}}^{-1}$ is a stellar damping time-scale, and $T_{\rm obs}$ is the total observing time. Here we combine this formula with a hierarchical Bayesian scheme to infer the population-level distribution of $n_{\rm pl}$ for a pulsar population of size $M$. The scheme is validated using synthetic data to quantify its accuracy systematically and prepare for its future application to real, astronomical data. For a plausible test population with $M=100$ and injected $n_{\rm pl}$ values drawn from a population-level Gaussian with mean $\mu_{\rm pl}=4$ and standard deviation $\sigma_{\rm pl}=0.5$, intermediate between electromagnetic braking and mass quadrupole gravitational radiation reaction, the Bayesian scheme infers $\mu_{\rm pl}=3.89^{+0.24}_{-0.23}$ and $\sigma_{\rm pl}=0.43^{+0.21}_{-0.14}$. The $M=100$ per-pulsar posteriors for $n_{\rm pl}$ and $\sigma^{2}_{\ddot{\nu}}\gamma_{\ddot{\nu}}^{-2}$ contain $87\%$ and $69\%$, respectively, of the injected values within their $90\%$ credible intervals. Comparable accuracy is achieved for (i) population sizes spanning the range $50 \leq M \leq 300$, with fractional errors ranging from $2\%$ to $6\%$ for $\mu_{\rm pl}$, and $12\%$ to $54\%$ for $\sigma_{\rm pl}$, and (ii) wide priors satisfying $\mu_{\rm pl} \leq 10^{3}$ and $\sigma_{\rm pl} \leq 10^{2}$, which accommodate plausible spin-down mechanisms with $\dot{K}\neq0$ and $\vert \dot{K} / K \vert \gg \vert \dot{\nu}/\nu\vert$. The Bayesian scheme generalizes readily to other plausible astrophysical situations, such as  pulsar populations with bimodal $n_{\rm pl}$ distributions. \\
\end{abstract}




\section{Introduction}
\label{Sec:Introduction}

The long-term evolution of the braking torque of a rotation-powered pulsar offers insights into the pulsar's magnetosphere and interior~\citep{BlandfordRomani1988}. It is studied through phase-coherent timing experiments by measuring the braking index,

\begin{equation}
    n=\frac{\nu \ddot{\nu}}{\dot{\nu}^{2}},
    \label{Eq:Intro_n_def}
\end{equation}

where $\nu$ is the pulse frequency, and the overdot symbolizes a derivative with respect to time. Plausible physical theories predict $\dot{\nu} = K \nu^{n_{\rm pl}}$, with $K$ constant and $n=n_{\rm pl}$ in the absence of stochastic fluctuations in $\nu$, e.g. due to instrumental factors, interstellar propagation or intrinsic (achromatic) noise. Physical examples of $\dot{\nu} \propto \nu^{n_{\rm pl}}$ include electromagnetic torques with $2 \lesssim n_{\rm pl} \leq 3$~\citep{GunnOstriker1969,Goldreich1970,Melatos1997,BucciantiniThompson2006,ContopoulosSpitkovsky2006,KouTong2015} or $n_{\rm pl} > 3$ when accounting for higher-order multipoles~\citep{Petri2015,Petri2017,AraujoDeLorenci2024}, as well as gravitational radiation reaction torques with $n_{\rm pl}=5$ (mass quadrupole, e.g. mountains)~\citep{Thorne1980}, and $n_{\rm pl}=7$ (current quadrupole, e.g. r-modes)~\citep{PapaloizouPringle1978,Andersson1998,OwenLindblom1998}.  

Phase-coherent timing experiments return measurements of $n$ consistent with an electromagnetic torque~\citep{LivingstoneKaspi2007,LivingstoneKaspi2011} for pulsars that are relatively free of intrinsic timing noise or rotational glitches. Examples include PSR J1640$-$4631 with $n=3.15\pm0.03$~\citep{ArchibaldGotthelf2016}, and PSR J0534$+$2200 with $n=2.51\pm0.01$~\citep{LyneAG1993}. However, for most rotation-powered pulsars where $n$ can be measured, timing experiments return $3 \ll \vert n \vert \lesssim 10^{6}$, with some pulsars exhibiting negative $n$ values~\citep{JohnstonGalloway1999,ChukwudeChidiOdo2016,LowerBailes2020,ParthasarathyJohnston2020,OnuchukwuLegahara2024}. High-$\vert n \vert$ values are termed `anomalous'. Population studies show that anomalous braking indices are correlated with glitch activity and timing noise~\citep{Cordes1980,ArzoumanianNice1994,JohnstonGalloway1999,UramaLink2006,LowerJohnston2021}. 

Several phenomenological modifications of $\dot{\nu}=K\nu^{n_{\rm pl}}$ have been proposed to explain anomalous braking indices~\citep{BlandfordRomani1988}. One alternative is that $K$ increases or decreases secularly on a time-scale which is short compared to the characteristic spin-down time-scale $\tau_{\rm sd} = \nu/(2\vert \dot{\nu} \vert)$, e.g.\ due to (counter-)alignment of the rotation and magnetic axes~\citep{Goldreich1970,LinkEpstein1997,Melatos2000,BarsukovPolyakova2009,JohnstonKarastergiou2017,AbolmasovBiryukov2024}, magnetic field evolution~\citep{TaurisKonar2001,PonsVigano2012}, Hall drift~\citep{BransgroveLevin2024}, precession~\citep{Melatos2000,BarsukovTsygan2010,WassermanCordes2022}, or magnetospheric switching~\citep{LyneHobbs2010,StairsLyne2019}. Another alternative, which receives special attention in this paper, is that the secular electromagnetic or gravitational torque ($2 \lesssim n_{\rm pl} \lesssim 7$) is masked by a stochastic torque, which dominates $\ddot{\nu}$ over typical observational timescales, e.g. due to relaxation processes mediated by crust-superfluid coupling~\citep{SedrakianCordes1998,AlparBaykal2006,GugercinogluAlpar2014,Gugercinoglu2017,LowerJohnston2021} or achromatic timing noise inherent to the stellar crust or superfluid core~\citep{CordesDowns1985,Jones1990,MelatosLink2014,ChukwudeChidiOdo2016}, as opposed to timing noise produced by propagation effects~\citep{GoncharovReardon2021}. Importantly, the two alternatives above are not mutually exclusive; they may coexist with each other~\citep{VargasMelatos2024}~and with other explanations~\citep{ChukwudeBaiden2010,ColesHobbs2011,OnuchukwuLegahara2024}.

Recently, \cite{VargasMelatos2023} calculated how a stochastic torque generates anomalous braking indices in the context of an idealized, phenomenological model, in which the fluctuations $\delta\ddot{\nu}(t)=\ddot{\nu}(t) - d [K \nu(t)^{n_{\rm pl}}]/dt$, with $\dot{K}=0$, obey standard, mean-reverting (i.e.\ damped) Brownian motion. That is, one has

\begin{equation}
    \frac{d[\delta\ddot{\nu}(t)]}{dt} = -\gamma_{\ddot{\nu}} \delta\ddot{\nu}(t) + \xi(t),
 \label{Eq:Intro_evolution_ddotnu}
\end{equation}

where $\gamma_{\ddot{\nu}}^{-1}$ is the mean-reversion time-scale, and $\xi(t)$ is a fluctuating, zero-mean, Langevin driver. \cite{VargasMelatos2023} derived (and verified through Monte-Carlo simulations with synthetic data) a predictive, falsifiable formula for the variance of the measured $n$, viz.

\begin{equation}
    \langle n^{2} \rangle = n_{\rm pl}^{2}+\frac{\sigma^{2}_{\ddot{\nu}}\nu^{2}}{\gamma_{\ddot{\nu}}^{2}\dot{\nu}^{4}T_{\rm obs}},
    \label{Eq:Intro_variance_n_MES}
\end{equation}

where $\sigma^2_{\ddot{\nu}} \propto \langle \xi(t) \xi(t') \rangle$ is the squared timing noise amplitude, $T_{\rm obs}$ is the total observing time, and angular brackets denote an average over an ensemble of random realizations of $\xi(t)$. Equation~(\ref{Eq:Intro_variance_n_MES}) can be applied to any pulsar with a measurement $n \sim n_{\rm pl} \pm \langle n^2 \rangle^{1/2}$ to infer statistically a relation between the parameters $\sigma^{2}_{\ddot{\nu}}/\gamma_{\ddot{\nu}}^{2}$ and $n_{\rm pl}$. The former parameter can be related to the phase residual power spectral density (PSD) $P_{\rm r}(f)$~\citep{LentatiAlexander2013,LentatiAlexander2014,GoncharovReardon2021,KeithNitu2023,VargasMelatos2023} and observational signatures of crust-superfluid coupling in the pulsar interior, such as glitch recovery time-scales or the autocorrelation time-scale of timing noise~\citep{PriceLink2012,MeyersMelatos2021,MeyersO'Neill2021,O'NeillMeyers2024}. As a starting point, \cite{VargasMelatos2023} applied equation (\ref{Eq:Intro_variance_n_MES}) to the representative pulsar PSR J0942$-$5552 observed by~\cite{LowerBailes2020}.  

In this paper we combine equation (\ref{Eq:Intro_variance_n_MES}) with a hierarchical Bayesian scheme to infer the distributions of the astrophysically important quantities $\sigma^{2}_{\ddot{\nu}}/\gamma^{2}_{\ddot{\nu}}$ and $n_{\rm pl}$ for a hypothetical sample of braking index measurements from $50 \leq M \leq 300$ synthetic pulsars. The analysis is done deliberately on synthetic data to validate the method and quantify its performance systematically under controlled conditions. The next step --- to apply the method to real astronomical data --- will be undertaken in collaboration with the pulsar timing community in a forthcoming paper. The paper is organized as follows. In Section~\ref{Sec: Methods (TBC)} we establish the inference scheme by embedding the Brownian model from~\cite{VargasMelatos2023} within a hierarchical Bayesian framework. In Section~\ref{Sec:Results (TBC)} we explain the validation procedure for the hierarchical Bayesian scheme. Section~\ref{subsec_III:results_M_100} uses the hierarchical Bayesian scheme to infer posterior distributions for $\sigma^{2}_{\ddot{\nu}}/\gamma_{\ddot{\nu}}^{2},$ and $n_{\rm pl}$ for a pulsar test population of size $M=100$, generated synthetically using the model described in Section~\ref{Sec: Methods (TBC)}. Section~\ref{subsec_III:varying_pop_size} verifies how the accuracy of the inference scheme varies over the plausible range $50 \leq M \leq 300$. In Section~\ref{sec:Koft} we generalize the analysis in Sections~\ref{subsec_III:results_M_100} and~\ref{subsec_III:varying_pop_size} to the scenario $\dot{K} \neq 0$ and $| \dot{K} / K | \gg | \dot{\nu} / \nu |$, motivated by some of the secular spin-down mechanisms cited in Section~\ref{Sec:Introduction}. In doing so, we test simultaneously the sensitivity of the results to the choice of prior, focusing on the situation where the prior is wide (i.e.\ permissive) and therefore conservative. The implications for future applications to real data are discussed briefly in Section~\ref{Sec:Conclusions}.  

\section{Hierarchical Bayesian scheme}
\label{Sec: Methods (TBC)}

 In this section, we explain how to construct a hierarchical Bayesian scheme, which combines pulsar timing data, a per-pulsar likelihood expressed in terms of per-pulsar parameters derived from equation (\ref{Eq:Intro_variance_n_MES}), and a population-level prior expressed in terms of population-level hyperparameters to calculate posterior distributions for $n_{\rm pl}$ and $\sigma_{\ddot{\nu}}^2 / \gamma_{\ddot{\nu}}^2$ individually for one pulsar and collectively for a population of $M$ pulsars. In Section~\ref{subsecII:Brownianmodel}, we introduce briefly the phenomenological model of rotational evolution which leads to equation (\ref{Eq:Intro_variance_n_MES}) \citep{VargasMelatos2023}, in order to clarify (i) what it means to measure $n$ in the left-hand side of equation (\ref{Eq:Intro_variance_n_MES}) from a sequence of pulse times of arrival (TOAs), and (ii) how to interpret the physical parameters $n_{\rm pl}$ and $\sigma_{\ddot{\nu}}^2/\gamma_{\ddot{\nu}}^2$ in the context of a pulsar timing experiment. The model in Section~\ref{subsecII:Brownianmodel} is also used to generate synthetic TOAs to test the hierarchical Bayesian scheme in Section \ref{Sec:Results (TBC)}. In Section~\ref{subsecII:Bayesframework}, we state the inference problem formally by specifying the mathematical form of the hierarchical version of Bayes's theorem which is the foundation of the scheme. The formalism in Section~\ref{subsecII:Bayesframework} is justified in terms of the general structure and definitions of a hierarchical model in Appendix~\ref{AppendixA:HB_model}. In Section~\ref{subsecII:chifromrms}, we specify the per-pulsar likelihood and the population-level prior and associated hyperparameters implemented in the Bayesian framework. Sections \ref{subsecII:Brownianmodel}--\ref{subsecII:chifromrms} present a complete, step-by-step procedure for applying the hierarchical scheme to real or synthetic data.
 
\subsection{Rotational evolution per pulsar} \label{subsecII:Brownianmodel}

A falsifiable, first-principles theory of the rotational evolution of a pulsar, including internal and magnetospheric torques, is not available at present \citep{AbolmasovBiryukov2024}. Instead, in this paper, we work with an idealized, phenomenological model, which seeks to approximate faithfully two observed properties which are relevant to measurements of anomalous braking indices, viz. secular spin down and stochastic timing noise (excluding glitches). Intuitively, the idea is that a pulsar spins down secularly in response to a torque $\propto \nu^{n_{\rm pl}}$, with $n_{\rm pl} \lesssim 7$ for many plausible mechanisms (see references in Section~\ref{Sec:Introduction}, neglecting $K$ evolution for now). It also executes mean-reverting Brownian motion about the secular trend in response to stochastic torques of uncertain origin (see references in Section~\ref{Sec:Introduction}). Mean reversion is important. Without it, an undamped random walk would grow in root-mean-square deviation and ``scramble'' the secular spin down over the star's lifetime, contradicting population synthesis studies~\citep{Faucher-GiguereKaspi2006} and the broken-power-law form of the phase residual PSD in many objects~\citep{GoncharovReardon2021,AntonelliBasu2022}. Physically plausible values of $\gamma_{\ddot{\nu}}$ and $\sigma_{\ddot{\nu}}$ in the Brownian model produce synthetic TOAs which resemble qualitatively the TOAs measured in real pulsars~\citep{VargasMelatos2023,VargasMelatos2024}, with fractional fluctuations much smaller than unity in $\nu$ and $\dot{\nu}$ but of order unity and greater in $\ddot{\nu}$. The latter property (along with $K$ evolution in general) contributes to anomalous braking indices satisfying $|n| \gg 1$. The fluctuations in $\ddot{\nu}$ and hence $n$ time-average to zero, if one observes for long enough, but ``long enough'' is longer than contemporary pulsar timing campaigns for many pulsars. Specifically, one requires $ T_{\rm obs} \gtrsim~3 \times10^{5} \linebreak \times(\sigma_{\ddot{\nu}}^{2} / 10^{-55} \, {\rm Hz}^2{\rm s}^{-5 }  ) (\gamma_{\ddot{\nu}}/10^{-6} \, {\rm s^{-1}})^{-2}(\dot{\nu} / 10^{-14} \, {\rm Hz \, s^{-1}})^{-4}(\nu / 1 \, {\rm Hz})^{2} \\
\,{\rm years}$, in order for the first term on the right-hand side of equation~(\ref{Eq:Intro_variance_n_MES}) to dominate the second term and deliver $n \approx n_{\rm pl}$~\citep{VargasMelatos2023,VargasMelatos2024}.

The instantaneous rotational state of a pulsar at time $t$ is described in general by the rotational phase $\phi(t)$ of the crust, the frequency $\nu(t)=\dot{\phi}(t)$, and its time derivatives $\dot{\nu}(t)$ and $\ddot{\nu}(t)$.\footnote{Third- and higher-order derivatives of $\nu(t)$ are part of the state description too, but they are not measured independently in most pulsars at present and are neglected in this paper.} These dynamical variables are packaged in the state vector ${\bf X}=(X_{1},X_{2},X_{3},X_{4})^{\rm T} = (\phi,\nu,\dot{\nu},\ddot{\nu})^{\rm T}$, where ${\rm T}$ denotes the matrix transpose. The state vector evolves according to the stochastic differential equation~\citep{MeyersMelatos2021,MeyersO'Neill2021,AntonelliBasu2022,VargasMelatos2023,O'NeillMeyers2024}

\begin{equation}
    d{\bf X}=({\bf A}{\bf X}+{\bf E})dt+{\bf \Sigma} d{\bf B}(t),
    \label{Eq_secII:dX}
\end{equation}

with

\begin{equation}
    \bm A = \begin{pmatrix} 0 & 1 & 0 & 0 \\ 0 & -\gamma_{\nu} & 1 & 0\\ 0 & 0 & -\gamma_{\dot{\nu}} & 1 \\
    0 & 0 & 0 & -\gamma_{\ddot{\nu}} \end{pmatrix}, \label{Eq_secII:Amplitudes_Matrix} 
\end{equation}

\begin{equation}
    \bm E = \begin{pmatrix} 0 \\ \gamma_{\nu} \nu_{\rm pl}(t) \\ \gamma_{\dot{\nu}} \dot{\nu}_{\rm pl}(t) \\ \dddot{\nu}_{\rm pl}(t)+\gamma_{\ddot{\nu}} \ddot{\nu}_{\rm pl}(t) \end{pmatrix},
    \label{Eq_secII:torque_vector_E}
\end{equation}

\noindent and

\begin{equation}
    \bm \Sigma = \text{diag}\left(0, 0, 0 ,\sigma_{\ddot{\nu}} \right). \label{Eq_secII:Sigma_Matrix}
\end{equation}

 The first two terms on the right-hand side of equation~(\ref{Eq_secII:dX}) are deterministic; they describe mean reversion to secular spin down. In equations~(\ref{Eq_secII:Amplitudes_Matrix}) and (\ref{Eq_secII:torque_vector_E}), the parameters $\gamma_{\nu},\gamma_{\dot{\nu}}$, and $\gamma_{\ddot{\nu}}$ are constant damping coefficients, $\nu_{\rm pl}(t)$ is the solution to the secular braking law $\dot{\nu}_{\rm pl} = K \nu_{\rm pl}^{n_{\rm pl}}$, and $\dot{\nu}_{\rm pl}(t),\ddot{\nu}_{\rm pl}(t),$ and $\dddot{\nu}_{\rm pl}(t)$ correspond to the first, second, and third time derivatives of $\nu_{\rm pl}(t)$, respectively. The third term on the right-hand side of equation~(\ref{Eq_secII:dX}) is stochastic; it describes a fluctuating Langevin driver, whose statistics are tuned to produce phase residuals consistent qualitatively with those observed in real pulsars, e.g.\ similar root-mean-square amplitude and autocorrelation time-scale. The parameter $\sigma^{2}_{\ddot{\nu}}$ in equation~(\ref{Eq_secII:Sigma_Matrix}) defines the amplitude of the Langevin driver, which we take to be a memoryless white-noise process, i.e. the Brownian increment $d{\bf B}(t)$ satisfies

\begin{equation}
    \langle dB_{i}(t)\rangle=0
    \label{Eq_SecII:dB_average}
\end{equation}

and

\begin{equation}
    \langle dB_{i}(t)dB_{j}(t') \rangle=\delta_{ij}\delta(t-t'),
    \label{Eq_SecII:dB_mem_less}
\end{equation} 

where $\langle ... \rangle$ denotes the average over an ensemble of random realizations of $d{\bf B}(t)$. For simplicity, we assume that there are no cross-correlations, viz. $\Sigma_{ij}=0$ for $i \neq j$ in equation~(\ref{Eq_secII:Sigma_Matrix}). It is easy to relax this assumption in the future if warranted when analyzing real data. We also assume $\Sigma_{11} = \Sigma_{22} = \Sigma_{33} = 0$ to ensure that the observables $\nu(t)$, $\dot{\nu}(t)$, and $\ddot{\nu}(t)$ are differentiable quantities; see Appendix A1 of \cite{VargasMelatos2023} for a discussion of this subtle issue. The model is modified easily to include fluctuations in other elements of ${\bf X}$, e.g. $\Sigma_{11} \neq 0$ describes noise in $\phi(t)$ due to magnetospheric fluctuations. The reader is encouraged to use different forms of ${\bf \Sigma}$ to suit the application at hand.

The Brownian model described by equations~(\ref{Eq_secII:dX})--(\ref{Eq_SecII:dB_mem_less}) can be solved analytically for ${\bf X}(t)$ given a specific random realization of $d{\bf B}(t)$; see Section 2.1 and Appendix A1 in \cite{VargasMelatos2023} for details. From the solution, one can calculate directly the braking index $n$ measured between the two epochs $t_{1}$ and $t_{2}>t_{1}$ according to the standard, nonlocal recipe~\citep{JohnstonGalloway1999}\footnote{A subtle distinction exists between the nonlocal measurement in equation (\ref{Eq_Apndx2:n_diffstates}) and the local measurement $n(t) = \nu(t) \ddot{\nu}(t) / \dot{\nu}(t)^2$ when handling a stochastic process theoretically; see Appendix A1 in \cite{VargasMelatos2023} for details.}

\begin{equation}
    n=1-\frac{\dot{\nu}(t_{1})\nu(t_{2})-\dot{\nu}(t_{2})\nu(t_{1})}{\dot{\nu}(t_{1})\dot{\nu}(t_{2})T_{\rm obs}}.
    \label{Eq_Apndx2:n_diffstates}
\end{equation}

Having calculated $n$ from equation~(\ref{Eq_Apndx2:n_diffstates}) for a single, random $d{\bf B}(t)$ realization, one averages over the $d{\bf B}(t)$ ensemble to obtain $\langle n \rangle = n_{\rm pl}$ and $\langle n^2 \rangle$ given by equation~(\ref{Eq:Intro_variance_n_MES}). Numerical solutions of equations~(\ref{Eq_secII:dX})--(\ref{Eq_SecII:dB_mem_less}) for astrophysically plausible values $\gamma_\nu \sim \gamma_{\dot{\nu}} \ll T_{\rm obs}^{-1} \ll \gamma_{\ddot{\nu}}$~\citep{PriceLink2012,MeyersMelatos2021,MeyersO'Neill2021,O'NeillMeyers2024} reproduce qualitatively the observed timing behavior of real pulsars, e.g. PSR J0942$-$5552~\citep{VargasMelatos2023,VargasMelatos2024}. Specifically, the zero-mean fluctuating variables $\delta \nu(t)=\nu(t) - \nu_{\rm pl}(t),~\delta \dot{\nu}(t)=\dot{\nu}(t) - \dot{\nu}_{\rm pl}(t)$, and $\delta \ddot{\nu}(t)=\ddot{\nu}(t) - \ddot{\nu}_{\rm pl}(t)$ satisfy $\vert \delta \nu(t) \vert \ll \vert \nu_{\rm pl}(t) \vert$, $\vert \delta \dot{\nu}(t) \vert \ll \vert \dot{\nu}_{\rm pl}(t) \vert$, and $\vert \delta \ddot{\nu}(t) \vert \gtrsim \vert \ddot{\nu}_{\rm pl}(t) \vert$ respectively.

\subsection{Statement of the inference problem} \label{subsecII:Bayesframework}

Consider a population of $M$ pulsars indexed by $1\leq m \leq M$. Each pulsar comes with a measurement $n^{(m)}_{\rm meas}$ of its braking index and an associated measurement uncertainty $\Delta n^{(m)}_{\rm meas}$. We obtain $n^{(m)}_{\rm meas}$ from the Taylor-expanded~\tempoDOS~ephemeris cubic fit by evaluating $\nu(t_{\rm ref}) \ddot{\nu}(t_{\rm ref})[\dot{\nu}(t_{\rm ref})]^{-2}$ at some reference time $0 \leq t_{\rm ref} \leq T_{\rm obs}$~\citep{ParthasarathyJohnston2020, LowerBailes2020}. Likewise, we evaluate $\Delta n^{(m)}_{\rm meas}=\nu(t_{\rm ref}) \Delta\ddot{\nu}(t_{\rm ref})[\dot{\nu}(t_{\rm ref})]^{-2}$, where $\Delta \ddot{\nu}(t_{\rm ref})$ is the nominal uncertainty for $\ddot{\nu}(t_{\rm ref})$ returned by the \tempoDOS~fitting process. \footnote{The measurement $\nu(t_{\rm ref}) \ddot{\nu}(t_{\rm ref})[ \dot{\nu}(t_{\rm ref})]^{-2}$ agrees with equation (\ref{Eq_Apndx2:n_diffstates}) up to negligible corrections of order $T_{\rm obs}$ divided by the spin-down time-scale; see also footnote 2.} Lists of $n^{(m)}_{\rm meas}$ measurements have been published by several authors for a total of $\sim 40$ pulsars at the time of writing, when one includes objects in the anomalous regime~\citep{LivingstoneKaspi2007,LivingstoneNg2011,WeltevredeJohnston2011,LyneJordan2015,LowerBailes2020,ParthasarathyJohnston2020,OnuchukwuLegahara2024}. Each pulsar also comes with an independent measurement of the root-mean-square of the timing residuals obtained after \tempoDOS's~cubic fit, $S_{\rm meas}^{(m)}$, and its associated uncertainty, $\Delta S_{\rm meas}^{(m)}$ obtained from the pulsar's TOA uncertainties. We explain the procedure to calculate $S_{\rm meas}^{(m)}$ and $\Delta S_{\rm meas}^{(m)}$ in Section~\ref{subsecIII:generate_synthetic_data}. In the general notation of a hierarchical Bayesian scheme defined in Appendix~\ref{AppendixA:HB_model}, the data for the $m$-th pulsar are denoted by $D^{(m)} = \{ n^{(m)}_{\rm meas} , \Delta n^{(m)}_{\rm meas} , S^{(m)}_{\rm meas}, \Delta S^{(m)}_{\rm meas}\}$. The data for all $M$ pulsars are denoted by $D = \{ D^{(1)}, \dots, D^{(M)} \}$.

Each pulsar is characterized by two parameters in the context of the Brownian model, namely $n_{\rm pl}^{(m)}$ and $\chi^{(m)}=\sigma_{\ddot{\nu}}^{(m)}[\gamma_{\ddot{\nu}}^{(m)}]^{-1}$.\footnote{The product $[\chi^{(m)} s^{(m)}]^{2}$, with $s^{(m)} = \nu^{(m)}[\dot{\nu}^{(m)}]^{-2}[\ T_{{\rm obs}}^{(m)}]^{-1/2}$, coincides with the right-most term in equation (\ref{Eq:Intro_variance_n_MES}). The bundle of observables $s^{(m)}$ can be measured with a precision of $1\%$ or better in principle, even though $T_{\rm obs}$ is not always reported to that precision in the literature. Hence $s^{(m)}$ is approximated as being known exactly a priori in this paper.} We seek to infer $n_{{\rm pl}}^{(m)}$ and $\chi^{(m)}$ for each pulsar individually while simultaneously inferring the $n_{{\rm pl}}^{(m)}$ distribution at a population level. The inference problem is hierarchical, because we assume that $n_{\rm pl}^{(m)}$ and $\chi^{(m)}$ are drawn from a single, universal, population-level prior distribution $\pi[n_{\rm pl}^{(m)}, \chi^{(m)} \vert \mu_{\rm pl}, \sigma_{\rm pl}]$, which takes the same form for all $M$ pulsars and is parameterized by two population-level hyperparameters, $\mu_{\rm pl}$ and $\sigma_{\rm pl}$. The hyperparameters are defined below and have their own prior distribution $\pi(\mu_{\rm pl}, \sigma_{\rm pl})$. That is, we do not apply per-pulsar priors $\pi[n_{\rm pl}^{(m)}, \chi^{(m)}]$ individually at the pulsar level; rather, we encode the prior information collectively at the population level through $\pi[n_{\rm pl}^{(m)}, \chi^{(m)} | \mu_{\rm pl}, \sigma_{\rm pl}]$ and $\pi(\mu_{\rm pl},\sigma_{\rm pl})$. In the general notation of a hierarchical model defined in Appendix~\ref{AppendixA:HB_model}, the per-pulsar parameters and population-level hyperparameters are denoted by $\theta^{(m)} = \{ n_{\rm pl}^{(m)} , \chi^{(m)} \}$ and $\psi= \{ \mu_{\rm pl}, \sigma_{\rm pl} \}$, respectively. The full set of parameters for all $M$ pulsars is denoted by $\theta= \{ \theta^{(1)}, \dots, \theta^{(M)} \}$.

The $(2M+2)$-dimensional joint posterior distribution $p(\psi, \theta | D)$ is calculated from the product of the likelihood and priors in the standard manner according to Bayes's theorem. Mathematically we write 

\begin{equation}
    p(\psi, \theta | D) = {\cal Z}^{-1}\prod_{m'=1}^{M} {\cal L}^{(m')}[D^{(m')} \vert \theta^{(m')}]\pi[\theta^{(m')} \vert \psi]\pi(\psi), 
    \label{eq_SecII:BAYES}
\end{equation}

with

\begin{equation}
    {\cal Z}=\int d\psi d\theta\prod_{m'=1}^{M} {\cal L}^{(m')}[D^{(m')} \vert \theta^{(m')}]\pi[\theta^{(m')} \vert \psi]\pi(\psi).
    \label{eq_SecII:Z_BAYES}
\end{equation}

Equations (\ref{eq_SecII:BAYES}) and (\ref{eq_SecII:Z_BAYES}) are justified in Appendix~\ref{AppendixA:HB_model} and follow from equations (\ref{Eq_appA:Bayestheorem_hyp})--(\ref{eq_appA:bayes_evidence_fact_exch}). The exchangeability of the $M$ independent pulsars permits the factorization of the joint posterior into the $M$-fold product in equation (\ref{eq_SecII:BAYES}), where ${\cal L}^{(m')}$ denotes the per-pulsar likelihood for the $m'$-th pulsar, defined in Section~\ref{subsecII:chifromrms}. The hierarchical assumption in the previous paragraph explains why the prior on $n_{\rm pl}^{(m')}$ and $\chi^{(m')}$ in the last line of equation (\ref{eq_SecII:BAYES}) is a product of the population-level prior $\pi[n_{{\rm pl}}^{(m')},\chi^{(m')} \vert \mu_{\rm pl}, \sigma_{\rm pl}]$ and the hyperparameter prior $\pi(\mu_{\rm pl},\sigma_{\rm pl})$, defined in Section~\ref{subsecII:chifromrms}.

We compute the joint posterior in equations (\ref{eq_SecII:BAYES}) and (\ref{eq_SecII:Z_BAYES}) using a Hamiltonian Monte Carlo No U-Turn Sampler \citep{Betancourt2017} implemented in \texttt{Stan} and run using the \texttt{cmdstanpy} interface. We then compute posterior distributions for any parameters of astrophysical interest by marginalizing over the other parameters in the usual way. In this paper, in preparation for analyzing real data, the main astrophysical motivation is to infer the population-level distribution of $n_{\rm pl}$, because this parameter can be related to the internal and magnetospheric physics of rotation-powered pulsars, as discussed in Sections~\ref{Sec:Introduction} and~\ref{subsecII:Brownianmodel}. With regard to $n_{\rm pl}$, the posterior distribution of the hyperparameters is given by 

\begin{equation} 
p(\mu_{\rm pl},\sigma_{\rm pl} | D) = \int d\theta \,p(\mu_{\rm pl}, \sigma_{\rm pl}, \theta | D),
\label{Eq_subsecII:marg_post_mu_sigma}
\end{equation} 

which corresponds to equations (\ref{eq_appA:A7})--(\ref{eq_appA:step_A9}) in Appendix~\ref{AppendixA:HB_model}. The population-level posterior distribution of $n_{\rm pl}$, renamed as $n_{\rm pl}^{({\rm pop})}$ to distinguish it from its per-pulsar counterpart $n_{\rm pl}^{(m)}$, is calculated from equation (\ref{Eq_subsecII:marg_post_mu_sigma}) according to 

\begin{align} 
p[n_{\rm pl}^{({\rm pop})} | D] =& \int d\mu_{\rm pl} \, d\sigma_{\rm pl} \, \pi[ n_{\rm pl}^{(m)} \mapsto n_{\rm pl}^{({\rm pop})} | \mu_{\rm pl}, \sigma_{\rm pl} ] \nonumber \\
&\times p(\mu_{\rm pl},\sigma_{\rm pl} | D), 
\label{Eq_subsecII:marg_post_n_pop}
\end{align}

which corresponds to equation (\ref{eq_appA:step13}) in Appendix~\ref{AppendixA:HB_model}. The notation $n_{\rm pl}^{(m)} \mapsto n_{\rm pl}^{({\rm pop})}$ in the first line of equation (\ref{Eq_subsecII:marg_post_n_pop}) means that the symbol $n_{\rm pl}^{(m)}$ is replaced by $n_{\rm pl}^{({\rm pop})}$ in the population-level prior.

\subsection{Per-pulsar likelihood and population-level prior} \label{subsecII:chifromrms}

The per-pulsar likelihood ${\cal L}^{(m)}[D^{(m)} | \theta^{(m)} ]$ in equations (\ref{eq_SecII:BAYES}) and (\ref{eq_SecII:Z_BAYES}) is the product of two factors, one related to measuring $n_{\rm meas}^{(m)}$, and the other related to measuring $S_{\rm meas}^{(m)}$. The first factor is approximately proportional to a Gaussian,

\begin{equation}
n_{\rm meas}^{(m)} \sim {\cal N}[ n_{\rm pl}^{(m)} , \{ [\chi^{(m)}s^{(m)}]^2 + [\Delta n_{\rm meas}^{(m)}]^2 \}^{1/2} ],
\label{Eq_secII:nmeas}
\end{equation}

where the notation $X \sim {\cal N}(a,b)$ symbolizes that the random variate $X$ is distributed as a Gaussian with mean $a$ and standard deviation $b$. Equation (\ref{Eq_secII:nmeas}) is essentially a restatement of equation (\ref{Eq:Intro_variance_n_MES}) including measurement errors, together with the property $\langle n \rangle = n_{\rm pl}$; see Appendix A1 in~\cite{VargasMelatos2023}. The Gaussian approximation follows from the linearity of the Brownian model in equations~(\ref{Eq_secII:dX})--(\ref{Eq_SecII:dB_mem_less}) and is validated through Monte Carlo simulations in previous work~\citep{VargasMelatos2023,VargasMelatos2024}. Empirically one finds $\Delta n_{\rm meas}^{(m)} \ll \chi^{(m)}s^{(m)}$ for most pulsars, consistent with \cite{VargasMelatos2023}, which simplifies the right-hand side of equation (\ref{Eq_secII:nmeas}) to ${\cal N}[n_{\rm pl}^{(m)}, \chi^{(m)}s^{(m)}]$ if necessary. The second factor in the per-pulsar likelihood is approximately proportional to a log-normal (base $e$),

\begin{equation}
\log S_{\rm meas}^{(m)} \sim {\cal N}[\mu_{ S \rm, BM}[\chi^{(m)}] , \{\sigma^{2}_{S \rm, BM} +[\Delta S_{\rm meas}^{(m)}]^{2}\}^{1/2} ].
\label{Eq_secII:zeta}
\end{equation}

Equation~(\ref{Eq_secII:zeta}) is justified formally in Appendix~\ref{AppendixB}, where the arguments of ${\cal N}(a,b)$ in equation~(\ref{Eq_secII:zeta}) are also defined. In summary,  ${\cal L}^{(m)}$ in equations (\ref{eq_SecII:BAYES}) and (\ref{eq_SecII:Z_BAYES}) is proportional to the product of the two distributions defined by equations (\ref{Eq_secII:nmeas}) and (\ref{Eq_secII:zeta}).

The population-level prior $\pi[ n_{\rm pl}^{(m)}, \chi^{(m)} | \mu_{\rm pl}, \sigma_{\rm pl} ]$ summarizes the existing astrophysical knowledge about the distribution of $n_{\rm pl}^{(m)}$ and $\chi^{(m)}$ in Milky Way pulsars. This knowledge is imperfect and debated~\citep{AbolmasovBiryukov2024}, so there are many valid ways to select $\pi[ n_{\rm pl}^{(m)}, \chi^{(m)} | \mu_{\rm pl}, \sigma_{\rm pl} ]$. In this paper, in preparation for analyzing real data, we assume that all $M$ pulsars spin down secularly via the same, low-$n_{\rm pl}$ mechanism, e.g.\ electromagnetic braking or gravitational radiation reaction (see Section~\ref{Sec:Introduction} and cf. Section~\ref{sec:Koft}). Under this assumption, $\pi[ n_{\rm pl}^{(m)}, \chi^{(m)} | \mu_{\rm pl}, \sigma_{\rm pl} ]$ is unimodal, with standard deviation $\sigma_{\rm pl}$ comparable to or less than $\mu_{\rm pl}$. In the absence of a compelling alternative, we set

\begin{equation}
    n_{{\rm pl}}^{(m)} \sim \mathcal{N}(\mu_{\rm pl}, \sigma_{\rm pl}).
    \label{Eq_secII:npl_distribution}
\end{equation}

We emphasize that equation (\ref{Eq_secII:npl_distribution}) tests a particular astrophysical scenario, which may not hold in reality. For example if some Milky Way pulsars are dominated by electromagnetic braking (with $n_{\rm pl}$ distributed narrowly about $n_{\rm pl}=3$) and others are dominated by gravitational radiation reaction (with $n_{\rm pl}$ distributed narrowly about $n_{\rm pl}=5$), equation (\ref{Eq_secII:npl_distribution}) should be replaced by a bimodal distribution. We also assume that there is no known astrophysical reason to prefer particular values of $\chi^{(m)}$ over others, so that $\pi[ n_{\rm pl}^{(m)}, \chi^{(m)} | \mu_{\rm pl}, \sigma_{\rm pl} ]$ is independent of $\chi^{(m)}$. That is, the population-level prior is approximately uniform and hence uninformative in $\chi^{(m)}$. This is also a reasonable approximation for two reasons. First, $\gamma_{\ddot{\nu}}$ in equation (\ref{Eq:Intro_variance_n_MES}) has not been measured in many pulsars, except for a few experiments studying the autocorrelation time-scale of pulsar timing noise~\citep{PriceLink2012,O'NeillMeyers2024}. \footnote{There is no guarantee that the damping mechanism governing $\gamma_{\ddot{\nu}}$ in the Brownian model in Section~\ref{Sec:Introduction} is the same as the damping mechanism governing glitch recoveries~\citep{GugercinogluAlpar2020,MelatosMillhouse2023}} Second, $\sigma_{\ddot{\nu}}$ in equation~(\ref{Eq:Intro_variance_n_MES}) is known to span many decades across the many Milky Way pulsars whose timing noise amplitudes have been measured, and there is no compelling preference for particular values~\citep{CordesHelfand1980,ColesHobbs2011,ParthasarathyShannon2019,LowerBailes2020,GoncharovReardon2021}. We accommodate the above properties through the loose restriction $\log\chi^{(m)} \sim \mathcal{N}(-19.5,5)$, to cover many decades of $\chi^{(m)}$ while ensuring numerical convergence. As the Bayesian scheme generalizes easily to other plausible astrophysical situations, we encourage the reader to try different forms of $\pi[ n_{\rm pl}^{(m)}, \chi^{(m)} | \mu_{\rm pl}, \sigma_{\rm pl} ]$, if the need arises.

The prior distribution of the hyperparameters $\mu_{\rm pl}$ and $\sigma_{\rm pl}$ is selected to be consistent with the astrophysical scenario tested in the previous paragraph, that all Milky Way pulsars spin down secularly via the same, low-$n_{\rm pl}$ mechanism taken from the plausible list in Section~\ref{Sec:Introduction} and references therein (cf. Section~\ref{sec:Koft}). In the absence of physical information to the contrary, we assume that $\mu_{\rm pl}$ and $\sigma_{\rm pl}$ are independent statistically, so that the hyperparameter prior factorizes as $\pi(\mu_{\rm pl},\sigma_{\rm pl}) = \pi(\mu_{\rm pl}) \pi(\sigma_{\rm pl})$. In what follows, we usually set $\mu_{\rm pl} \sim \mathcal{U}(2, 8)$, where $X\sim{\cal U}(a,b)$ denotes that $X$ is distributed uniformly on the domain $a \leq X \leq b$. Alternative priors are valid, of course, and can be substituted easily by the reader. We also set $\log \sigma_{\rm pl}\sim \mathcal{N}\left(0, 1\right)$ and verify a posteriori that the results are insensitive to this choice. A wider set of priors is also analyzed in the sensitivity test in Section~\ref{sec:Koft}. Finally, to assist with the computation, we truncate arbitrarily the Gaussian in equation (\ref{Eq_secII:npl_distribution}) to the same domain $2 \leq n_{\rm pl}^{(m)} \leq 8$ as the $\mu_{\rm pl}$ prior. Truncated thus,  equation (\ref{Eq_secII:npl_distribution}) has zero weight at $n_{\rm pl}=0$, assisting with the convergence of the sampling algorithm which evaluates equation~(\ref{eq_SecII:BAYES}). 
 
\begin{table}
\centering
\caption{Prior ranges used by the hierarchical Bayesian analysis in Sections~\ref{subsec_III:results_M_100} and~\ref{subsec_III:varying_pop_size} for the population-level hyperparameters $\psi$ (upper half) and the per-pulsar parameters $\theta^{(m)}$ (lower half). The last two columns define ${\cal U}(a,b)$ or ${\cal N}(a,b)$ according to the prior used (third column). The reader is encouraged to experiment with different priors, depending on the application at hand.}
\label{Table_subsecII:priorsHB}
\begin{tabular*}{\columnwidth}{p{0.15\columnwidth}p{0.15\columnwidth}p{0.2\columnwidth}p{0.15\columnwidth}p{0.15\columnwidth}}
\hline
Parameter & Units & Prior & $a$ & $b$ \\
\hline
 $\mu_{\rm pl}$ & --- & Uniform & 2 & 8 \\
 $\sigma_{\rm pl}$ & --- & Log-normal & 0 & 1\\
 \hline 
 $n_{\rm pl}^{(m)}$ & --- & Normal & $\mu_{\rm pl}$ & $\sigma_{\rm pl}$ \\
 $\chi^{(m)}$ & s$^{-5/2}$ & Log-normal & -19.5 & 5 \\
\hline
\end{tabular*}
\end{table}

Table~\ref{Table_subsecII:priorsHB} summarizes the priors used in Sections~\ref{subsec_III:results_M_100} and~\ref{subsec_III:varying_pop_size} for the population-level hyperparameters $\psi$ and the per-pulsar parameters $\theta^{(m)}$.

\section{Setting up the validation tests}
\label{Sec:Results (TBC)}

Validation tests involving synthetic data are an important prelude to astronomical applications involving real data. They are a tool to quantify systematically the accuracy with which physical parameters can be inferred under controlled conditions, where their injected values are known. To this end, this section sets up the main steps in the validation procedure, namely the recipe to generate the data, $D$, ingested by the Bayesian scheme, and how we quantify the accuracy of the inference scheme. The section is organized as follows. Section~\ref{subsecIII:syntheticTOAs} briefly explains how to generate synthetic TOAs from the phenomenological model in Section~\ref{subsecII:Brownianmodel}. Section~\ref{subsecIII:generate_synthetic_data} explains how to convert the TOAs into per-pulsar synthetic measurements of $n_{\rm meas}^{(m)}$, $\Delta n_{\rm meas}^{(m)}$, $S_{\rm meas}^{(m)}$, and $\Delta S_{\rm meas}^{(m)}$, which are packaged as $D$. Section~\ref{subsecIII:error} defines the fractional error metric used to quantify the accuracy of the inference scheme.

\subsection{Synthetic TOAs} \label{subsecIII:syntheticTOAs}

We generate synthetic TOAs for each of the $M$ pulsars in the test population by solving equations~(\ref{Eq_secII:dX})--(\ref{Eq_SecII:dB_mem_less}) numerically for a specific realization of $d{\bf B}(t)$ with a randomly selected seed. Each numerical realization of the Brownian model ingests the model parameters $\{\sigma^{2}_{\ddot{\nu}}, \gamma_{\nu},\gamma_{\dot{\nu}}, \gamma_{\ddot{\nu}}\}
$, the pulsar's right ascension (RA) and declination (DEC), and the initial conditions ${\bf X}(t_{0})$ to produce a set of $N_{\rm TOA}$ synthetic TOAs  alongside their uncertainty $\Delta_{\rm TOA}$. For simplicity, we assume that $\Delta_{\rm TOA}$ is the same for every TOA, although it is easy to relax this assumption if the need arises. In practice, the TOAs are generated using a sample of times $t_{i}$ satisfying $X_{1}(t_{i})=\phi(t_{i})~{\rm mod}~2\pi =0$, within the interval $0 \leq t_{i} \leq T_{\rm obs}$. Every realization of the Brownian model with a new random seed generates unique, yet statistically equivalent, ${\bf X}(t_{i})$ and hence TOA sequences. The recipe to generate the $t_{i}$ samples is described in full in Section~2.2 of \cite{VargasMelatos2023}.\footnote{The procedure to generate synthetic data is implemented in the publicly available {\tt baboo} package at \url{http://www.github.com/meyers-academic/baboo}.}

\subsection{Synthetic per-pulsar measurements $D^{(m)}$} \label{subsecIII:generate_synthetic_data}

To generate $D=\{D^{(1)},\dots,D^{(M)}\}$, we randomly select $M$ pulsars from the Australian Telescope National Facility (ATNF) Pulsar Catalogue~\citep{ManchesterHobbs2005} and record their ${\rm RA},{\rm DEC},\nu$, and $\dot{\nu}$ values, such that the $m$-th pulsar is parameterized by $\{{\rm RA}^{(m)},{\rm DEC}^{(m)},\nu^{(m)},\dot{\nu}^{(m)}\}$ for $1 \leq m \leq M$. For the sake of definiteness, we choose pulsars in the ATNF catalogue which satisfy $1 \leq \nu/(1~{\rm Hz}) \leq 10$. We equip each pulsar with a random $n_{\rm pl}^{(m)}$ value drawn from equation~(\ref{Eq_secII:npl_distribution}) with $\mu_{\rm pl}=4$ and $\sigma_{\rm pl}=0.5$.\footnote{The choice $\mu_{\rm pl}=4$ has no particular physical significance. It is a reasonable intermediate value for testing purposes, at the midpoint between electromagnetic braking ($n=3$) and mass quadrupole gravitational radiation reaction ($n=5$).} We also compute $\ddot{\nu}^{(m)}=n_{\rm pl}^{(m)}[\nu^{(m)}]^{-1}[\dot{\nu}^{(m)}]^{2}$, from equation~(\ref{Eq:Intro_n_def}). The values of $\nu^{(m)},\dot{\nu}^{(m)}$, and $\ddot{\nu}^{(m)}$ are packaged into the per-pulsar initial condition vector ${\bm X}^{(m)}(t_{0})=[0,\nu^{(m)},\dot{\nu}^{(m)},\ddot{\nu}^{(m)}]^{\rm T}$, where we set the initial phase to be $\phi^{(m)}=0$ without loss of generality. The parameters $T_{\rm obs}^{(m)}$ and $N_{\rm TOA}^{(m)}$ are drawn the from uniform distributions $T_{\rm obs}^{(m)}/(1~{\rm year}) \sim {\cal U}(7,15)$ and $N_{\rm TOA}^{(m)} \sim {\cal U}(80,300)$, respectively, representative of current pulsar timing campaigns~\citep{ParthasarathyShannon2019, LowerBailes2020}. For the Brownian model parameters $\{\sigma^{2}_{\ddot{\nu}}, \gamma_{\nu},\gamma_{\dot{\nu}}, \gamma_{\ddot{\nu}}\}
$, we draw $\log_{10}\{ [\sigma^{(m)}_{\ddot{\nu}}]^{2} /(1~{\rm Hz}^{2}{\rm s}^{-5})\} \sim {\cal U}(-55,-48)$, $\log_{10}[\gamma_{\ddot{\nu}}^{(m)}/(1~{\rm s}^{-1})] \sim {\cal U}(-7,-5)$, and fix $\gamma_{\nu}=\gamma_{\dot{\nu}}=10^{-13}~{\rm s}^{-1}$. The foregoing range of $\sigma^{2}_{\ddot{\nu}}$ typically yields anomalous braking indices~\citep{VargasMelatos2023,VargasMelatos2024}. The range of the damping coefficients guarantees that the synthetic timing residuals qualitatively resemble those observed in real pulsars,  satisfying $\gamma_\nu \sim \gamma_{\dot{\nu}} \ll [T_{\rm obs}^{(m)}]^{-1} \ll \gamma_{\ddot{\nu}}^{(m)}$, as discussed in Appendix A of \cite{VargasMelatos2023}. The priors for $[\sigma^{(m)}]^{2}$ and $[\gamma_{\ddot{\nu}}^{(m)}]^{2}$ imply that $\chi^{(m)}$ spans the range $(10^{-45}~{\rm s}^{-5}, 10^{-34}~{\rm s}^{-5})$. Table~\ref{Table_subsecII:priors_data} summarizes the distribution of the injected parameters used to generate $D^{(m)}$.

\begin{table}
\centering
\caption{Distribution of the injected parameters used to generate $D^{(m)}$ for the tests in Sections~\ref{subsec_III:results_M_100} and~\ref{subsec_III:varying_pop_size}. The last two columns define ${\cal U}(a,b)$ or ${\cal N}(a,b)$ according to the prior used (third column).}
\label{Table_subsecII:priors_data}
\begin{tabular*}{\columnwidth}{p{0.2\columnwidth}p{0.1\columnwidth}p{0.2\columnwidth}p{0.15\columnwidth}p{0.15\columnwidth}}
\hline
Parameter & units & Prior & $a$ & $b$ \\
\hline
 $n_{\rm pl}^{(m)}$ & --- & Normal & 4 & 0.5 \\
 $T_{\rm obs}^{(m)}$ & year & Uniform & 7 & 15 \\
 $N_{\rm TOA}^{(m)}$ & --- & Uniform & 80 & 300 \\
 $[\sigma_{\ddot{\nu}}^{(m)}]^{2}$ & ${\rm Hz}^{2}$\;s$^{-5}$ & Log-Uniform & $-55$ & $-48$ \\
 $\gamma_{\ddot{\nu}}^{(m)}$ & s$^{-1}$ & Log-Uniform & $-7$ & $-5$ \\
\hline
\end{tabular*}
\end{table}

We generate synthetic TOAs $\{t_{1}^{(m)},\dots,t_{N_{\rm TOA}^{(m)}}^{(m)}\}$ for the $m$-th pulsar by inputting the parameters $\{{\rm RA}^{(m)},{\rm DEC}^{(m)}, {\bf X}^{(m)}(t_{0}), T_{\rm obs}^{(m)}, N_{\rm TOA}^{(m)}, [\sigma^{(m)}_{\ddot{\nu}}]^{2}, \gamma_{\nu}, \gamma_{\dot{\nu}}, \gamma_{\ddot{\nu}}^{(m)}\}$ into the recipe detailed in Section~\ref{subsecIII:syntheticTOAs}. The latter step generates a set of TOA samples, $\{t_{i}^{(m)}\}$, for all $M$ pulsars. The TOAs are analyzed using~\tempoDOS~to generate a timing ephemeris, which includes the measured values $\nu^{(m)}(t_{\rm ref}), \dot{\nu}^{(m)}(t_{\rm ref}),\ddot{\nu}^{(m)}(t_{\rm ref})$, and $\Delta \ddot{\nu}^{(m)}(t_{\rm ref})$ for $0 \leq t_{\rm ref} \leq T_{\rm obs}^{(m)}$, the measured timing residuals $\{ {\cal R}[t_1^{(m)}], \dots, {\cal R}[ t_{N_{\rm TOA}^{(m)}}^{(m)} ] \}$, and the associated uncertainties $\{ \Delta{\cal R}[t_1^{(m)}], \dots, \Delta{\cal R}[ t_{N_{\rm TOA}^{(m)}}^{(m)} ] \}$. We calculate $n_{\rm meas}^{(m)}$ and $\Delta n_{\rm meas}^{(m)}$ from the \tempoDOS~fit as specified in Section~\ref{subsecII:Bayesframework}. We calculate $S^{(m)}_{\rm meas}$ from the timing residuals according to

\begin{equation}
    S^{(m)}_{\rm meas} = \sqrt{\frac{1}{N_{\rm TOA}^{(m)}}\sum_{i=1}^{N_{\rm TOA}^{(m)}} {\cal R}[t_{i}^{(m)}]^{2} }.\label{eq_subsecIII:S_meas_m}
\end{equation}

The associated uncertainty, $\Delta S^{(m)}_{\rm meas}$, is obtained via a bootstrapping method which assumes that the $i$-th timing residual, with $1 \leq i \leq N_{\rm TOA}^{(m)}$, is drawn from ${\cal N}\{{\cal R}[t_{i}^{(m)}], \Delta{\cal R}[t_{i}^{(m)}]\}$. From the latter normal distribution, we generate $10^{5}$ new ${\cal R}[t_{i}^{(m)}]$ samples for ${1 \leq i \leq N_{\rm TOA}^{(m)}}$, and combine these through Equation~(\ref{eq_subsecIII:S_meas_m}). The standard deviation of the $10^{5}$ new $S_{\rm meas}^{(m)}$ samples is $\Delta S_{\rm meas}^{(m)}$. 

From $n_{\rm meas}^{(m)}, \Delta n_{\rm meas}^{(m)}, S^{(m)}_{\rm meas},$ and $\Delta S_{\rm meas}^{(m)}$ we construct $D^{(m)}$ for $1\leq m \leq M$. To orient the reader, Table~\ref{Table_subsecII:example_synthDms} presents five examples of the resulting synthetic per-pulsar measurements, $D^{(m)}$, generated for five random realizations.

\begin{table}
\centering
\caption{Examples of $D^{(m)}$ generated following the recipe in Section~\ref{subsecIII:generate_synthetic_data} for five synthetic random realizations.}
\label{Table_subsecII:example_synthDms}
\begin{tabular*}{\columnwidth}{c p{0.175\columnwidth}p{0.175\columnwidth}p{0.175\columnwidth}p{0.175\columnwidth}}
\hline
$m$ & $n_{\rm meas}$ & $\Delta n_{\rm meas}$ & $S_{\rm meas}/(1~{\rm s})$ & $\Delta S_{\rm meas}/(1~{\rm s})$  \\
\hline
1 & $-7.46$ $\times$ 10$^1$ & 1.93 $\times$ 10$^{-2}$ &  5.39 $\times$ 10$^{-2}$ & 7.08 $\times$ 10$^{-7}$ \\
2 & 6.47 $\times$ 10$^2$ & 3.89 $\times$ 10$^0$ &  1.87 $\times$ 10$^{-4}$ & 3.24 $\times$ 10$^{-9}$ \\
3 & 2.48 $\times$ 10$^0$ & 1.65 $\times$ 10$^{-1}$ &  1.20 $\times$ 10$^{-4}$ & 2.03 $\times$ 10$^{-9}$ \\
4 & 1.25 $\times$ 10$^1$ & 3.13 $\times$ 10$^{-4}$ &  3.24 $\times$ 10$^{-2}$ & 4.50 $\times$ 10$^{-7}$ \\
5 & 6.00 $\times$ 10$^3$ & 6.83 $\times$ 10$^{-3}$ &  2.49 $\times$ 10$^0$ & 4.58 $\times$ 10$^{-5}$ \\
\hline
\end{tabular*}
\end{table}

\subsection{Quantifying the accuracy of the inference scheme} \label{subsecIII:error}

We quantify the accuracy of the inference scheme in two ways: a first-pass point estimate based on the posterior median, and a fuller treatment based on percentile-percentile (PP) plots.

With regard to the first-pass point estimate, we define the fractional error between an injected parameter value, $\theta_{\rm inj}'$, and the median of the corresponding one-dimensional posterior, as

\begin{equation}
    {\rm ERR}(\theta')=\frac{ \Big \vert \underset{\theta'}{\rm med}[p(\theta' \vert D)]-\theta_{\rm inj}' \Big \vert}{\vert \theta_{\rm inj}' \vert}.
    \label{eq_subsecIII:error_theta}
\end{equation}

In equation~(\ref{eq_subsecIII:error_theta}), $p(\theta' \vert D)$ is the one-dimensional posterior for a single parameter $\theta' \in \{ \mu_{\rm pl}, \sigma_{\rm pl}\}\cup \{n_{\rm pl}^{(m)}, \chi^{(m)} \}_{1\leq m \leq M}$ obtained from the Monte Carlo sampler, i.e. $p(\psi, \theta \vert D)$ marginalized over all the other  per-pulsar parameters and population-level hyperparameters except $\theta'$ [see equation~(\ref{eq_appA:posterior_mth_pulsar})]. We note that ${\rm ERR}(\theta')$ quantifies the closeness of the median of $\theta'$ to $\theta_{\rm inj}'$. It does not quantify if the one-dimensional posterior $p(\theta' \vert D)$ contains $\theta_{\rm inj}'$  within its 90\% credible interval, for example. We discuss the latter property in tandem in Section~\ref{subsec_III:results_M_100}, for a test population with $M=100$, and in Section~\ref{subsec_III:varying_pop_size}, for test populations with $M=50$ and $M=300$.

To construct a PP plot~\citep{Cook2006}, we analyze 100 realizations of $D$ generated according to Section~\ref{subsecIII:generate_synthetic_data}. For each realization we compute $p(\theta' \vert D)$ for $\theta' \in \{ \mu_{\rm pl}, \sigma_{\rm pl}\}\cup \{n_{\rm pl}^{(m)}, \chi^{(m)} \}_{1\leq m \leq M}$. The PP plot compares the fraction of $\theta'_{\rm inj}$ realizations included within a given credible interval of $p(\theta' \vert D)$, against the credible interval used. The Bayesian scheme perfectly recovers the $\theta'$ parameter if the PP plot is a diagonal line of unit slope. A PP plot displays more of the information contained in $p(\theta' \vert D)$, beyond simply its median, and is more informative than ${\rm ERR}(\theta')$. We present PP plots in Section~\ref{subsec_III:varying_pop_size} for test populations with $M=50,100,$ and $300$.

\section{Inference output for a test population with $M=100$} \label{subsec_III:results_M_100}

We start the validation exercise by applying the procedure in Section~\ref{subsecIII:generate_synthetic_data} to a sample of $M=100$ pulsars. The sample is comparable in size to other samples analyzed previously in studies of anomalous braking indices~\citep{ParthasarathyJohnston2020}. We fit the synthetic data to the model described in Section~\ref{subsecII:Bayesframework}. The Hamiltonian Monte Carlo No U-Turn Sampler~\citep{Betancourt2017} returns samples from the posterior $p(\psi, \theta\vert D)$ [equation~(\ref{eq_SecII:BAYES})].

\subsection{Population-level accuracy} \label{subsecIV:pop_lev_acc}
\begin{figure}
\flushleft
 \includegraphics[width=\columnwidth]{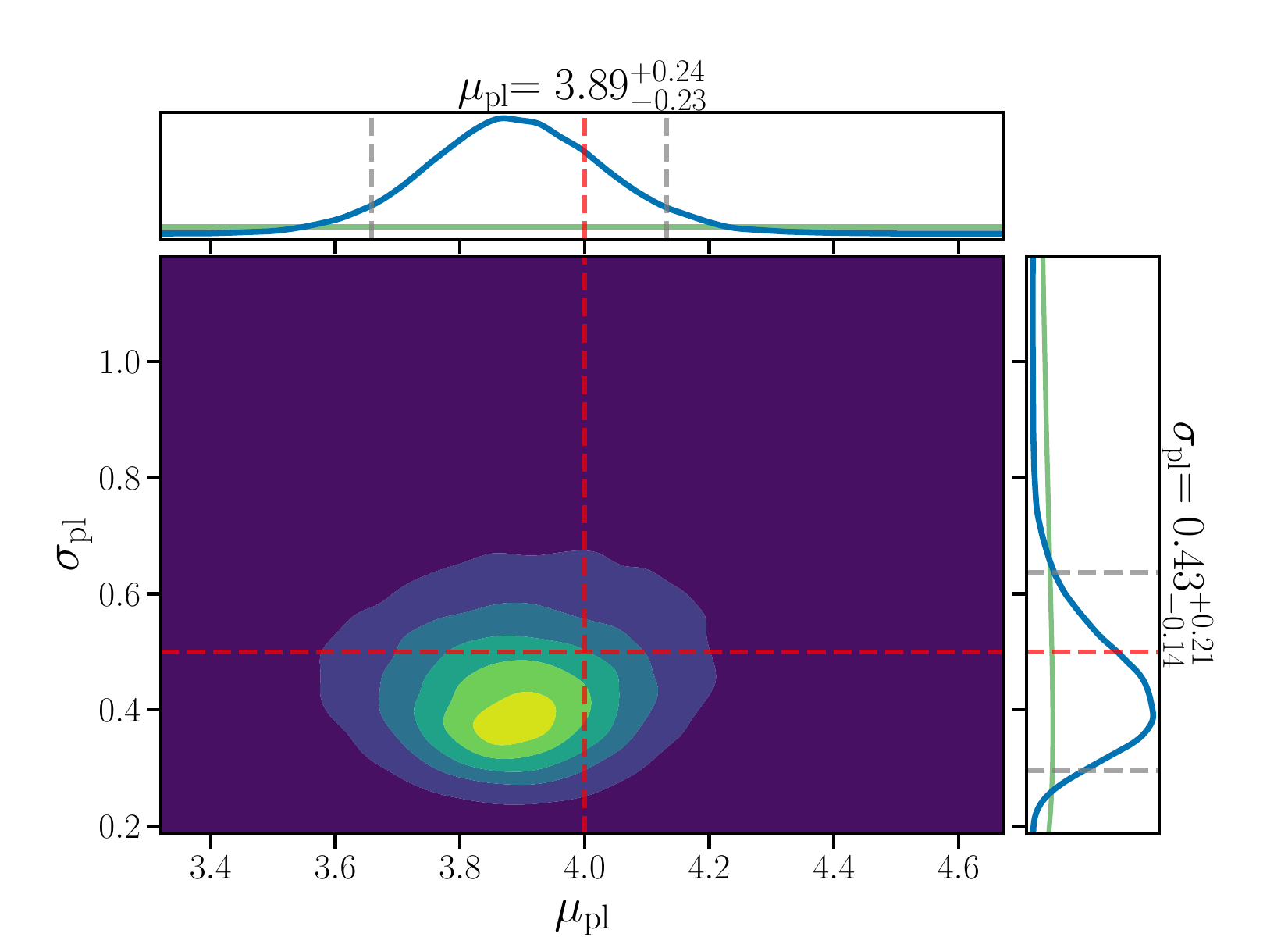}
 \caption{Posterior distribution $p(\mu_{\rm pl}, \sigma_{\rm pl} \vert D)$ of the population-level braking index hyperparameters $\mu_{\rm pl}$ and $\sigma_{\rm pl}$ for a population size of $M=100$ (Section~\ref{subsec_III:results_M_100}). The vertical and horizontal dashed red lines indicate the injected values $\mu_{\rm pl}=4$ and $\sigma_{\rm pl}=0.5$, respectively. The contours in the two-dimensional posterior (central panel) mark the $(30\%, 50\%, 70\%, 90\%)$-credible intervals. The one-dimensional posteriors correspond to $p(\psi, \theta \vert D)$ marginalized over every parameter except $\mu_{\rm pl}$ (upper panel) and $\sigma_{\rm pl}$ (right panel). In both the upper and right panels, the marginalized one-dimensional posteriors are plotted as blue curves, while the priors are plotted as green curves.  The supertitles of the upper and right panels quote the corresponding inferred median (central value) and $90\%$ credible intervals (error bars, indicated by grey dashed lines). The inferred medians of $\mu_{\rm pl}$ and $\sigma_{\rm pl}$ are displaced from the injected values by 2.8 and 14 per cent, respectively [see equation~(\ref{eq_subsecIII:error_theta})]. The injected values of $\mu_{\rm pl}$ and $\sigma_{\rm pl}$ fall within the inferred $90\%$ credible intervals.}
\label{fig_subsecII:corner_plot_100_pulsars}
\end{figure}

Figure~\ref{fig_subsecII:corner_plot_100_pulsars} presents the posterior distribution $p(\mu_{\rm pl}, \sigma_{\rm pl}\vert D)$ [equation~(\ref{Eq_subsecII:marg_post_mu_sigma})] in the form of a standard corner plot. The central panel displays the two-dimensional posterior $p(\mu_{\rm pl}, \sigma_{\rm pl}\vert D)$, where the contours mark the $(30\%,50\%,70\%,90\%)$-credible intervals. The central estimates of $\mu_{\rm pl}$ (upper panel) and $\sigma_{\rm pl}$ (right panel) are reported at the top of the respective one-dimensional posteriors. The central value corresponds to the posterior median, while the errors bars define a $90\%$ credible interval, bracketed by two grey vertical dashed lines in the one-dimensional posteriors. The injected values for the hyperparameters, $\mu_{\rm pl}=4$ and $\sigma_{\rm pl}=0.5$, are marked by red, dashed lines in the one-dimensional posteriors and in the contour plot for the two-dimensional posterior. In the upper and right panels, the green lines represent the priors for $\mu_{\rm pl}$ and $\sigma_{\rm pl}$, respectively.  From equation~(\ref{eq_subsecIII:error_theta}), we obtain ${\rm ERR}(\mu_{\rm pl})=2.8\times10^{-2}$ and ${\rm ERR}(\sigma_{\rm pl})=1.4\times10^{-1}$. The injected values for $\mu_{\rm pl}$ and $\sigma_{\rm pl}$ lie within the $90\%$ credible intervals. 

\begin{figure}
\flushleft
 \includegraphics[width=\columnwidth]{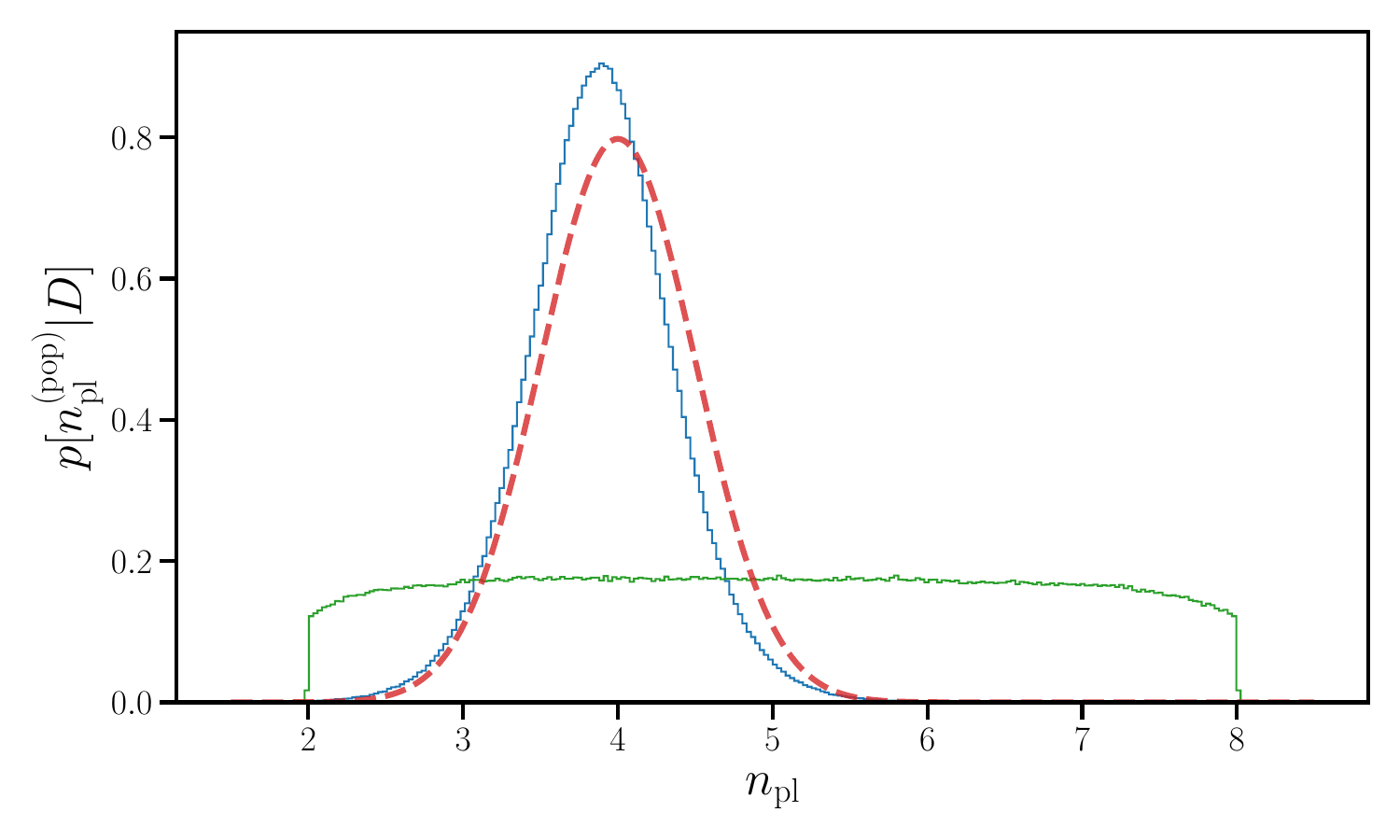}
 \caption{Population-level posterior distribution derived from Figure~\ref{fig_subsecII:corner_plot_100_pulsars} in the form of a posterior predictive check, i.e. a comparison between what the hierarchical Bayesian scheme infers and the injected braking index distribution. The inferred posterior distribution $p[n^{(\rm pop)}_{\rm pl} \vert D]$ (blue histogram) is calculated by combining $p(\mu_{\rm pl}, \sigma_{\rm pl} \vert D)$ (central panel of Figure~\ref{fig_subsecII:corner_plot_100_pulsars}) with the population level prior $\pi[n_{\rm pl}^{(m)} \mapsto n_{\rm pl}^{({\rm pop})} \vert \mu_{\rm pl}, \sigma_{\rm pl}]$ via equation~(\ref{Eq_subsecII:marg_post_n_pop}). The injected $n_{\rm pl}^{({\rm pop})}$ distribution is plotted as a red, dashed curve, while the prior $\pi[n_{\rm pl}^{({\rm pop})}]$ [equation~(\ref{Eq_secII:npl_distribution})] is plotted as the green histogram. The significant overlap between the blue histogram and red, dashed curve indicate that the hierarchical Bayesian scheme is able to recover accurately the braking index distribution at a population level with $M=100$. Both the blue histogram and the red, dashed curve follow approximately a Gaussian distribution.}
\label{fig_subsecII:ppc_100_pulsars}
\end{figure}

Another, equivalent way to visualize the results in Figure~\ref{fig_subsecII:corner_plot_100_pulsars} is to perform a posterior predictive check. Specifically, we construct the posterior distribution $p[n^{\rm (pop)}_{\rm pl} \vert D]$ in Figure~\ref{fig_subsecII:ppc_100_pulsars} (blue histogram) from $p(\mu_{\rm pl}, \sigma_{\rm pl} \vert D)$ in Figure~\ref{fig_subsecII:corner_plot_100_pulsars} via equation~(\ref{Eq_subsecII:marg_post_n_pop}).  The posterior $p[n_{\rm pl}^{({\rm pop})} | D]$ represents the population-level distribution of braking indices predicted by the inference scheme, when the pulsars are treated as indistinguishable a posteriori, i.e.\ by polling the $M$ pulsars collectively without attaching individual braking indices to individual pulsars. By way of comparison, we plot the injected $p[n^{\rm (pop)}_{\rm pl} \vert D]$ as the red dashed curve, and the prior $\pi[n_{\rm pl}^{\rm (pop)} \vert \mu_{\rm pl}, \sigma_{\rm pl}]$ [equation~(\ref{Eq_secII:npl_distribution}); see Section~\ref{subsecII:chifromrms}] as the green histogram. The recovered and injected $p[n^{\rm (pop)}_{\rm pl} \vert D]$ distributions overlap significantly, indicating that the Bayesian scheme recovers $n_{\rm pl}$ accurately at a population level. 

\begin{figure}
\flushleft
 \includegraphics[width=\columnwidth]{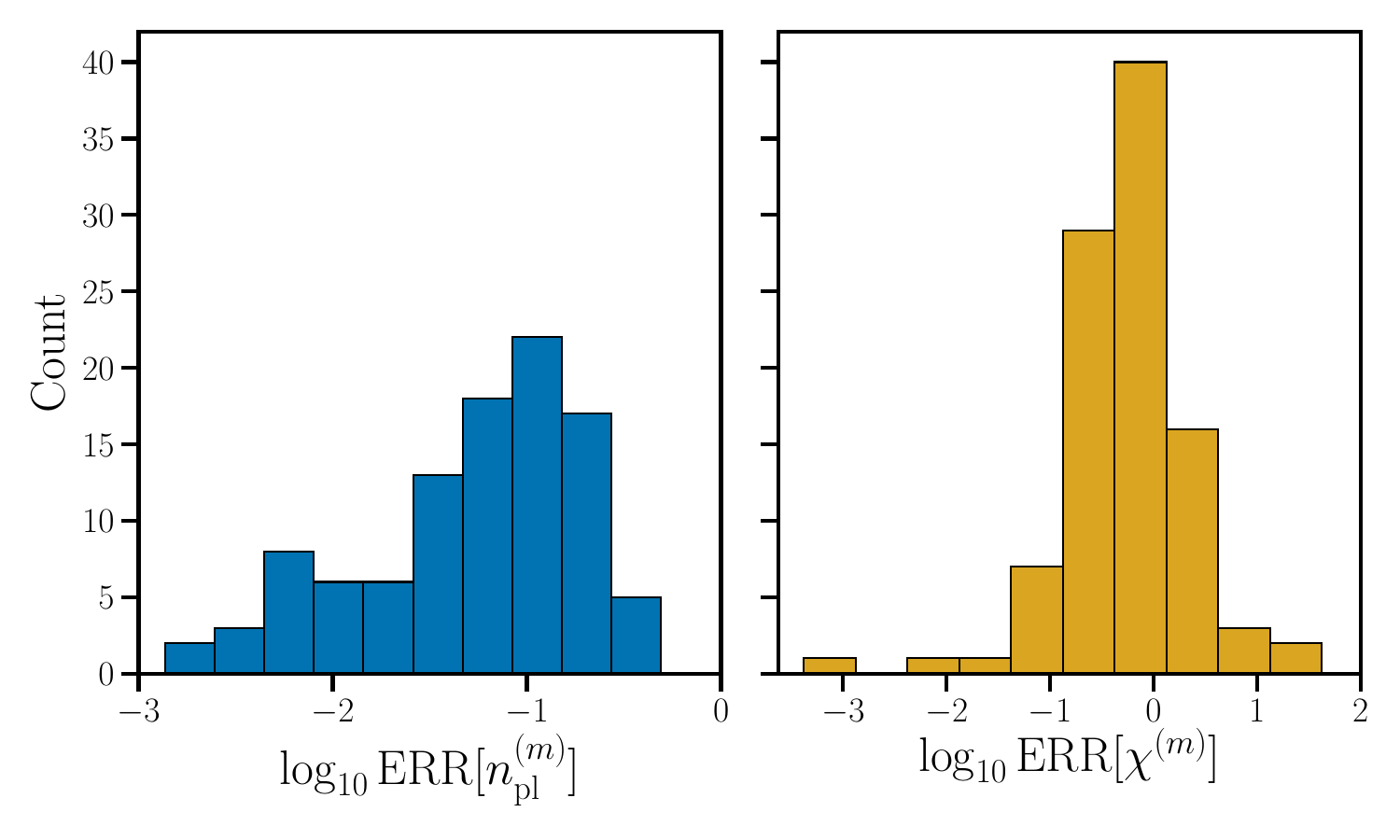}
 \caption{Per-pulsar accuracy of the inference scheme: histograms of the fractional error ${\rm ERR}(\theta')$, defined by~(\ref{eq_subsecIII:error_theta}), between the injected values and the median of the one-dimensional posteriors marginalized over all parameters except $\theta'=n_{\rm pl}^{(m)}$ (blue histogram, left panel) and $\theta'=\chi^{(m)}$ (orange histogram, right panel), with $1 \leq m \leq 100$, for the pulsars in the test population in Section~\ref{subsec_III:results_M_100} and Figures~\ref{fig_subsecII:corner_plot_100_pulsars} and~\ref{fig_subsecII:ppc_100_pulsars}. The horizontal axes have a base-10 logarithmic scale, with bin widths of $0.26$ dex (blue histogram) and $0.5$ dex (orange histogram). The hierarchical Bayesian scheme estimates $n_{\rm pl}^{(m)}$ more accurately than $\chi^{(m)}$ with ${\rm ERR}\Big[ n^{(m)}_{\rm pl} \Big]=9.6\times10^{-2}$ and ${\rm ERR}\Big[ \chi^{(m)} \Big]=1.7$ on average across the 100 pulsars.} 
\label{fig_subsecII:error_panels}
\end{figure}

\subsection{Per-pulsar accuracy} \label{subsec_IV:per_pulsar_Accura}

The hierarchical Bayesian scheme returns a posterior $p[\theta^{(m)} | D]$ for the parameters $\theta^{(m)} = \{ n_{\rm pl}^{(m)}, \chi^{(m)} \}$ of the $m$-th pulsar, calculated from equation~(\ref{eq_appA:posterior_mth_pulsar}). It is interesting to ask how accurately the scheme infers $\theta^{(m)}$ for every pulsar individually. For example, does the scheme accurately infer $\theta^{(m)}$ for most $m$, consistent with its good performance at the population level in Figures~\ref{fig_subsecII:corner_plot_100_pulsars} and~\ref{fig_subsecII:ppc_100_pulsars}, while also returning inaccurate $\theta^{(m)}$ estimates for a few $m$ values? Or does it infer $\theta^{(m)}$ with comparable accuracy for all $m$? 

To help answer these questions, Figure~\ref{fig_subsecII:error_panels} presents the histograms for ${\rm ERR}\Big[n_{\rm pl}^{(m)}\Big]$ (left panel, blue histogram) and ${\rm ERR}\Big[ \chi^{(m)} \Big]$ (right panel, orange histogram) for $1 \leq m \leq 100$, i.e. the fractional error between the inferred (median) and injected values for $n_{\rm pl}$ and $\chi$ polled across all $100$ pulsars in the test population. The blue histogram spans a range of $\sim 2.6~{\rm dex}$, with minimum and maximum errors of ${\rm min}\Big\{{\rm ERR}\Big[n_{\rm pl}^{(m)}\Big]\Big\}=1.4\times10^{-3}$ and ${\rm max}\Big\{{\rm ERR}\Big[n_{\rm pl}^{(m)}\Big]\Big\}=0.48$, respectively. The Bayesian scheme successfully disentangles the secular braking index $n_{\rm pl}^{(m)}$ from the anomalous $n_{\rm meas}^{(m)}$ values, which span $0 \leq \log_{10} \vert n_{\rm meas}^{(m)} \vert  \leq 10$. For instance, the pulsar corresponding to ${\rm max}\Big\{{\rm ERR}\Big[n_{\rm pl}^{(m)}\Big]\Big\}$, labelled $m=15$, has an injected value of $n_{\rm pl}^{(m=15)}=2.6$ and an estimated value of $n_{\rm pl}^{(m=15)}=3.9$. In comparison, for this pulsar we measure~$n_{\rm meas}^{(m=15)}=1.10\times10^{2}$, squarely within the anomalous regime. For 87 out of the 100 pulsars, the $90\%$ credible interval for the one-dimensional posterior $p\Big[n^{(m)}_{\rm pl} \vert D \Big]$ contains the injected $n^{(m)}_{\rm pl}$. In contrast, the $\chi^{(m)}$ accuracy is worse. The orange histogram in Figure~\ref{fig_subsecII:error_panels} spans $\approx 5~{\rm dex}$, with ${\rm min}\Big\{{\rm ERR}\Big[\chi^{(m)} \Big]\Big\}=4.2\times10^{-3}$ and ${\rm max}\Big\{{\rm ERR}\Big[\chi^{(m)} \Big]\Big\}=42$. This is not surprising, for two reasons. First, the prior domain for $n_{\rm pl}^{(m)}$, i.e. $2 \leq n_{\rm pl}^{(m)} \leq8$, is narrower than for $\chi^{(m)}$, which covers $\approx 30$ dex. Second, as discussed in Appendix~\ref{AppendixB}, the distribution of $S_{\rm meas}^{(m)}$ values consistent with a fixed $\chi^{(m)}$ value is broad, with $\sigma_{S \rm,BM} \approx 0.52$ [see equation~(\ref{Eq_secII:zeta})]. We summarize the minimum and maximum fractional $n^{(m)}_{\rm pl}$ and $\chi^{(m)}$ errors returned by this test in Table~\ref{Table:max_min_errs_Ms}. 

For 69 out of the 100 pulsars, the $90\%$ credible interval for the one dimensional posterior $p\Big[\chi^{(m)} \vert D \Big]$ contains the injected $\chi^{(m)}$. For the other 31 out of 100 pulsars, the inference scheme consistently overestimates $\chi^{(m)}$. The latter $31$ pulsars populate the tails of the blue and orange histograms in Figure~\ref{fig_subsecII:error_panels}. They can be divided into two categories. Category I includes $29$ pulsars which satisfy $\vert n^{(m)}_{\rm meas}-n^{(m)}_{\rm pl} \vert \gtrsim \chi^{(m)}_{\rm inj}s^{(m)}$, i.e. the measured $n^{(m)}_{\rm meas}$ is displaced by at least one standard deviation $\chi^{(m)}_{\rm inj}s^{(m)}$ from the injected $n_{\rm pl}^{(m)}$ value. Five out of these $29$ pulsars are within the non-anomalous braking index regime, with $\chi^{(m)}_{\rm inj}s^{(m)} \leq 0.5$~\citep{VargasMelatos2023}. For these five pulsars, we find $n^{(m)}_{\rm meas} \approx n_{\rm pl}^{(m)}$ with ${\rm max}\Big\{{\rm ERR}\Big[n^{(m)}_{\rm pl} \Big]\Big\}\lesssim 7.6\times10^{-2}$, so the degree of overestimation is small and does not matter in practice; non-anomalous braking indices are measured accurately regardless. The remaining $24$ pulsars are within the anomalous braking index regime with $ 1 \leq \chi^{(m)}_{\rm inj}s^{(m)} \leq 10^{8}$. For these $24$ pulsars, the measured $n_{\rm meas}^{(m)}$ value lies in the tail [$n_{\rm meas}^{(m)} \gtrsim \chi^{(m)}_{\rm inj}s^{(m)}$] of the distribution of possible $n_{\rm meas}^{(m)}$ at a fixed $\chi_{\rm inj}^{(m)}$ value [see Figure~2 of \cite{VargasMelatos2023}], so that the Bayesian scheme overestimates $\chi^{(m)}$. Category II includes two pulsars which satisfy $\vert n^{(m)}_{\rm meas}-n^{(m)}_{\rm pl} \vert \leq \chi^{(m)}_{\rm inj}s^{(m)}$, i.e. the measured $n_{\rm meas}^{(m)}$ is displaced by less than one standard deviation from the injected $n_{\rm pl}^{(m)}$. 

\section{Population size} \label{subsec_III:varying_pop_size}

The test population with $M=100$ analyzed in Section~\ref{subsec_III:results_M_100} matches approximately the number of pulsars with measured (non-)anomalous braking indices analyzed in previous studies~\citep{ParthasarathyJohnston2020,LowerBailes2020}. However, before applying the inference scheme in Section~\ref{Sec: Methods (TBC)} to real astronomical data, it is important to test how the accuracy of the inference output depends on $M$. Specifically, it is important for the sake of efficiency to determine the minimum $M$, for which results of tolerable accuracy are achieved, as well as the extent to which future measurements of additional braking indices will yield improvements in accuracy. To these ends, we generate test populations with $M=50$ and $M=300$ and apply the procedure in Section~\ref{Sec:Results (TBC)} to  repeat the validation experiment in Section~\ref{subsec_III:results_M_100}.

\begin{figure*}
\flushleft
 \includegraphics[width=2\columnwidth]{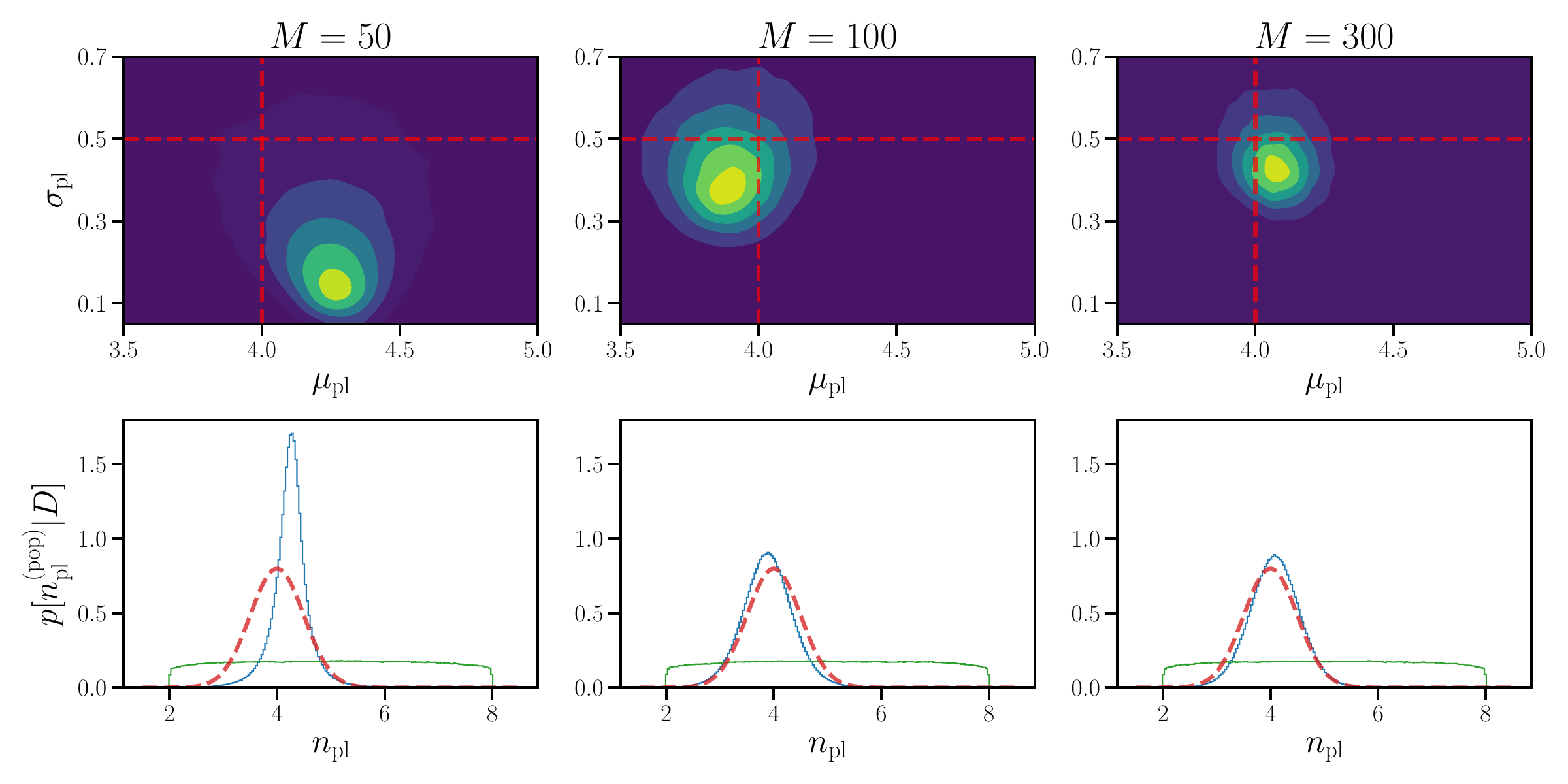}
 \caption{Effect of sample size: key posterior information produced by the hierarchical Bayesian scheme for test populations with $M=50$ (left column), $M=100$ (central column; copied from Figures~\ref{fig_subsecII:corner_plot_100_pulsars} and~\ref{fig_subsecII:ppc_100_pulsars}), and $M=300$ (right column). The top row displays contours of the two-dimensional posterior for the hyperparameters $p(\mu_{\rm pl},\sigma_{\rm pl}\vert D)$, with the credible intervals and injected values for $\mu_{\rm pl}$ and $\sigma_{\rm pl}$ color-coded as in Figure~\ref{fig_subsecII:corner_plot_100_pulsars}. The bottom row displays $p[n^{({\rm pop})}_{\rm pl} \vert D]$ as a posterior predictive check (same presentation as Figure~\ref{fig_subsecII:ppc_100_pulsars}). The accuracy of the inferred $\mu_{\rm pl}$ and $\sigma_{\rm pl}$ values increases with $M$.}
\label{fig_subsecII:ppc_various_Ms}
\end{figure*}

Figure~\ref{fig_subsecII:ppc_various_Ms}  presents for comparison the key posterior information produced by the validation tests for $M=50$ (left column), $M=100$ (central column;  copied from Figures~\ref{fig_subsecII:corner_plot_100_pulsars} and~\ref{fig_subsecII:ppc_100_pulsars}), and $M=300$ (right column). The top row displays contours of the two-dimensional posterior for the hyperparameters, $p(\mu_{\rm pl},\sigma_{\rm pl} \vert D)$, color-coded by credible intervals in the same fashion as the central panel in Figure~\ref{fig_subsecII:corner_plot_100_pulsars}. The bottom row displays $p[n_{\rm pl}^{({\rm pop})}\vert D]$ as a posterior predictive check, copying the presentation in Figure~\ref{fig_subsecII:ppc_100_pulsars}. The blue histograms in the bottom row represent the inferred $n_{\rm pl}^{({\rm pop})}$, the red, dashed curves represent the injected $p[n^{\rm (pop)}_{\rm pl} \vert D]$, and the green histograms represents the population-level prior $\pi[n_{\rm pl}^{\rm (pop)} \vert \mu_{\rm pl}, \sigma_{\rm pl}]$, as defined in Section~\ref{subsecII:chifromrms}. For $M=50,100,$ and $300$, we obtain $\mu_{\rm pl}=4.25^{+0.24}_{-0.29}$ and $\sigma_{\rm pl}=0.23^{+0.39}_{-0.13}$, $\mu_{\rm pl}=3.89^{+0.24}_{-0.23}$ and $\sigma_{\rm pl}=0.43^{+0.21}_{-0.14}$, and  $\mu_{\rm pl}=4.07^{+0.16}_{-0.17}$ and $\sigma_{\rm pl}=0.44^{+0.15}_{-0.10}$, respectively. The central value corresponds to the one-dimensional posterior median, and the error bars define the $90\%$ credible interval. Encouragingly, the injected $\mu_{\rm pl}$ and $\sigma_{\rm pl}$ are contained within the 90\% credible interval for $50\leq M \leq 300$. Equation~(\ref{eq_subsecIII:error_theta}) yields fractional errors of ${\rm ERR}(\mu_{\rm pl})=6.3\times10^{-2}~(M=50)$, $2.8\times10^{-2}~(M=100)$, and $2.0\times10^{-2}~(M=300)$, as well as ${\rm ERR}(\sigma_{\rm pl})=5.4\times10^{-1}~(M=50),1.4\times10^{-1}~(M=100)$, and $1.2\times10^{-1}~(M=300)$. The accuracy increases broadly with $M$, as expected. However, it is tolerable even for $M=50$, if $\mu_{\rm pl}$ is the primary quantity of physical interest.   

On a per-pulsar basis, the minimum and maximum fractional $n_{\rm pl}^{(m)}$ and $\chi^{(m)}$ errors for $50 \leq M \leq 300$ are summarized in Table~\ref{Table:max_min_errs_Ms}. For $M=50$, $66\%$ of the pulsars return $p[n^{(m)}_{\rm pl} \vert D]$ posteriors which include the injected $n^{(m)}_{\rm pl}$ value within the $90\%$ credible intervals, while for $M=100$ and $M=300$ this percentage is $87\%$. Taking $50 \leq M \leq 300$ together, $70\%$ of the pulsars return $p[\chi^{(m)}\vert  D]$ posteriors which include the injected $\chi^{(m)}$ value within the $90\%$ credible interval. The other $30\%$ do not fall within the $90\%$ credible interval and belong to categories I and II in Section~\ref{subsec_III:results_M_100}.  The maximum error does not decrease monotonically with $M$, because the chance of an outlier increases, as $M$ increases.

\begin{table}
\caption{Minimum and maximum fractional $n_{\rm pl}^{(m)}$ and $\chi^{(m)}$ errors for test populations of size $M=50, 100$, and $300$ in Section~\ref{subsec_III:varying_pop_size}. The subscript $m$ on ${\rm min}_{m}$ and ${\rm max}_{m}$ indicates that the extremum is taken with respect to the $M$ objects labeled by $1 \leq m \leq M$. The fractional errors ${\rm ERR}\Big[n_{\rm pl}^{(m)}\Big]$ and ${\rm ERR}\Big[\chi^{(m)}\Big]$ span 2.8 dex and 4.9 dex, 2.6 dex and 5.0 dex, and 3.5 dex and 4.3 dex, for $M=50,100$, and $300$, respectively.}
\centering
\begin{tabularx}{\columnwidth}{XccXcX}
\hline
& \multicolumn{2}{c}{${\rm ERR}\Big[n_{\rm pl}^{(m)}\Big]$} & \multicolumn{2}{c}{${\rm ERR}\Big[\chi^{(m)}\Big]$ } \\
$M$ & ${\rm min}_{m}$ & ${\rm max}_{m}$ & ${\rm min}_{m}$ & ${\rm max}_{m}$ \\
\hline
50 & $1.3\times10^{-2}$ & $8.2\times10^{-1}$ & $7.8\times10^{-2}$ & $7.5\times10^{2}$  \\
100 & $1.4\times10^{-3}$ & $4.8\times10^{-1}$ & $4.2\times10^{-3}$ & $4.2\times10^{1}$   \\
300 & $2.3\times10^{-4}$ & $7.8\times10^{-1}$ & $8.8\times10^{-3}$ & $1.8\times10^{2}$  \\
\hline
\end{tabularx}
\label{Table:max_min_errs_Ms}
\end{table}

\begin{figure*}
\centering
 \includegraphics[width=1.95\columnwidth]{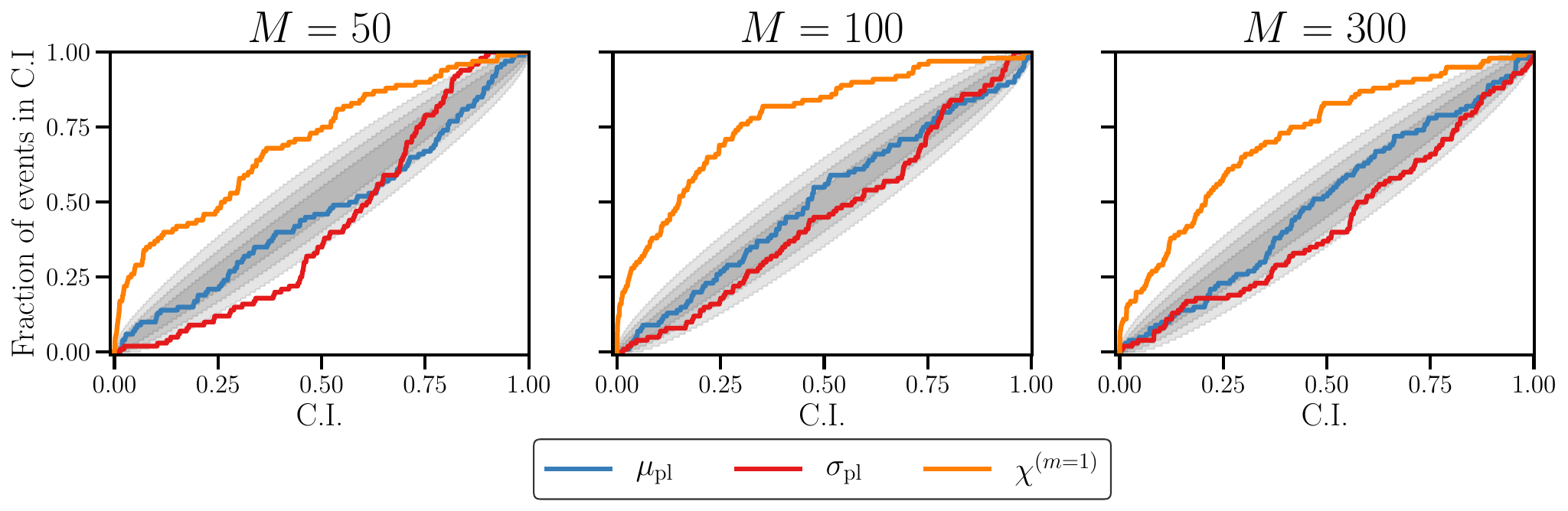}
 \caption{PP plots: fraction of injections within a given credible interval of $p(\theta' \vert D)$ (vertical axis) for $\theta'=\mu_{\rm pl}, \sigma_{\rm pl},$ and $\chi^{(m=1)}$ (see legend), versus the credible interval itself (horizontal axis) for test populations with $M=50$ (left panel), $M=100$ (central panel), and $M=300$ (right panel). Each curve is constructed from 100 realizations of $D$, generated following Section~\ref{subsecIII:generate_synthetic_data}. The shaded grey contours denote the $(1,2,3)$-$\sigma$ significance levels. A parameter is estimated accurately, if its corresponding PP curve approaches closely the diagonal of unit slope. Overall, the accuracy of the inference scheme improves, as $M$ increases, consistent with the results in Figure~\ref{fig_subsecII:ppc_various_Ms}. The accuracy is highest and lowest for $\mu_{\rm pl}$ and $\chi^{(m)}$ respectively.}
\label{fig_subsecII:PP_plots}
\end{figure*}

Figure~\ref{fig_subsecII:PP_plots} displays PP plots for the parameters $\mu_{\rm pl},\sigma_{\rm pl},$ and $\chi^{(m=1)}$ for $M=50$ (left panel), $M=100$ (central panel), and $M=300$ (right panel). The pulsar $m=1$ is representative; the other $M-1$ objects return similar PP plots for $\chi^{(m)}$ but are not drawn to avoid congestion. The shaded grey contours mark the $(1,2,3)$-$\sigma$ significance levels for $100$ realizations. We observe three features. First, the PP curves for $\mu_{\rm pl}$ and $\sigma_{\rm pl}$ lie fully within the 3-$\sigma$ envelope for $M \geq 100$ and mostly within the 3-$\sigma$ envelope for $M\geq 50$, whereas the PP curves for $\chi^{(m)}$ lie mostly outside the 3-$\sigma$ envelope for $M\geq 50$. Second, the estimates of all three parameters improve, as $M$ increases, in the sense that the PP curves approach the diagonal with unit slope more closely, as $M$ increases. Third, the bow-shaped PP curves for $\chi^{(m=1)}$ (and other $m$ values not plotted) indicate, that lower-value credible intervals contain more realizations, i.e.\ $p[\chi^{(m)} | D]$ is broad for all $m$. This is consistent with the results reported for ${\rm ERR} \left[ \chi^{(m)} \right]$ in Section~\ref{subsec_IV:per_pulsar_Accura} and Table~\ref{Table:max_min_errs_Ms}. The hierarchical Bayesian scheme infers $\mu_{\rm pl}$ and $\sigma_{\rm pl}$ more accurately than $\chi^{(m)}$ under a wide set of circumstances, because the constraints on $\chi^{(m)}$ derived from the data as well as prior information are relatively loose.

\section{Sensitivity to the prior for $\dot{K}\neq 0$} \label{sec:Koft}

The results in Sections~\ref{subsec_III:results_M_100} and~\ref{subsec_III:varying_pop_size} are obtained with reference to a torque law with $\dot{K} = 0$, relatively low $n_{\rm pl} \sim {\cal N}(\mu_{\rm pl}=4,\sigma_{\rm pl}=0.5)$, and correspondingly scaled priors. It is natural to ask two related questions in response. (i) Is the output of the hierarchical Bayesian scheme sensitive to the chosen priors, e.g.\ does the scheme infer $\mu_{\rm pl}$ and $\sigma_{\rm pl}$ accurately, even when the priors in Table~\ref{Table_subsecII:priorsHB} are widened substantially? (ii) Does the scheme continue to perform accurately when it is extended to a torque law with $\dot{K} \neq 0$ and $| \dot{K} / K | \gg | \dot{\nu} / \nu |$? \footnote{Sometimes the situation in question (ii) is interpreted informally as $\dot{K}=0$ and $\vert n_{\rm pl}\vert \gg 1$, but the conditions $\dot{K} \neq 0$ and $| \dot{K} / K | \gg | \dot{\nu} / \nu |$ are more plausible physically for most of the braking mechanisms cited in Section~\ref{Sec:Introduction}.} With respect to question (ii), the reader is reminded that a plausible alternative to explain anomalous braking indices is that $K$ evolves on a time-scale $\tau_{K}$ (much) shorter than the pulsar's spin-down time-scale $\tau_{\rm sd}$ (see Section~\ref{Sec:Introduction}). For this alternative, \cite{VargasMelatos2024} showed that one measures $\vert n_{\rm meas}\vert \approx \vert n_{\rm pl}+\dot{K}_{\rm dim}\vert \gg n_{\rm pl}$, with $\dot{K}_{\rm dim}$ given by

\begin{equation}
    \dot{K}_{\rm dim} = (n_{\rm pl}-1)\Big(1-\frac{K_{2}}{K_{1}}\Big)\frac{\tau_{\rm sd}}{\tau_{K}},
    \label{Eq:dot_K_dim}
\end{equation} 

with $K_{1} = K(t=0)$ and $K_{2} = K(t\rightarrow\infty)$. Importantly, this secular alternative may coexist with a stochastic torque ($\sigma_{\ddot{\nu}}^{2} \neq 0$), which also contributes to $\vert n_{\rm meas} \vert \gg n_{\rm pl}$. In this section, we address questions (i) and (ii) simultaneously by extending the validation tests in Sections~\ref{subsec_III:results_M_100} and~\ref{subsec_III:varying_pop_size} to the scenario $\dot{K} \neq 0$ while adopting relatively wide (i.e.\ permissive) and therefore conservative priors.

The hierarchical Bayesian scheme generalizes readily to the case $\dot{K} \neq 0$ and $\sigma_{\ddot{\nu}}^{2} \neq 0$. One simply replaces $n_{\rm pl}$ everywhere with $n_{\rm pl}+\dot{K}_{\rm dim}$.\footnote{There is no way to disentangle $n_{\rm pl}$ and $\dot{K}_{\rm dim}$ from their sum, given time-averaged data of the form $D^{(m)}=\{n^{(m)}_{\rm meas} , \Delta n^{(m)}_{\rm meas} , S^{(m)}_{\rm meas}, \Delta S^{(m)}_{\rm meas}\}$, using either the Bayesian scheme in this paper or other schemes in the literature.} \cite{VargasMelatos2024} proved analytically that the variance $\langle n^{2} \rangle$ [equation (\ref{Eq:Intro_variance_n_MES})] is modified according to

\begin{equation}
    \langle n^{2} \rangle = (n_{\rm pl}+\dot{K}_{\rm dim})^{2}+\frac{\sigma^{2}_{\ddot{\nu}}\nu^{2}}{\gamma_{\ddot{\nu}}^{2}\dot{\nu}^{4}T_{\rm obs}},
    \label{eq:new_variance_n_koft}
\end{equation}

which in turn, modifies the per-pulsar likelihood for $n_{\rm meas}^{(m)}$ [equation (\ref{Eq_secII:nmeas})] according to

\begin{equation}
n_{\rm meas}^{(m)} \sim {\cal N}[ (n_{\rm pl}+\dot{K}_{\rm dim})^{(m)} , \{ [\chi^{(m)}s^{(m)}]^2 + [\Delta n_{\rm meas}^{(m)}]^2 \}^{1/2} ].
\label{Eq_secII:nmeas_Koft}
\end{equation}

Equation~(\ref{eq:new_variance_n_koft}) makes it plain, that the only mathematical change in the inference problem is to replace $n_{\rm pl}$ by $n_{\rm pl}+\dot{K}_{\rm dim}$ everywhere, i.e. testing the performance of the hierarchical Bayesian scheme does not involve changing the algorithm in any way. The only changes involved are (i) to generate synthetic data with $\dot{K} \neq 0$ [see \cite{VargasMelatos2024}], and (ii) to widen the priors of $\mu_{\rm pl}$ and $\sigma_{\rm pl}$ (see Table~\ref{Table_subsecII:priorsHB}) to accommodate the wider range of $n_{\rm pl}+\dot{K}_{\rm dim}$ implied by $\dot{K}\neq0$.

By way of validation, we repeat the experiment detailed in Section~\ref{subsec_III:results_M_100} for a sample of $M=100$ synthetic pulsars assembled according to the recipe in Section~\ref{subsecIII:generate_synthetic_data}, except that now the effective braking indices $(n_{\rm pl}+\dot{K}_{\rm dim})^{(m)}$ are drawn from the wider distribution ${\cal N}(\mu_{\rm pl}=2,\sigma_{\rm pl}=50)$, as opposed to ${\cal N}(\mu_{\rm pl}=4,\sigma_{\rm pl}=0.5)$. This choice guarantees $\vert (n_{\rm pl}+\dot{K}_{\rm dim})^{(m)} \vert \gg 1$ for many $m$. For the random sample we analyze, we find the range $-128 \leq (n_{\rm pl}+\dot{K}_{\rm dim})^{(m)} \leq 100$. The Bayesian scheme is executed as detailed in Section~\ref{subsec_III:results_M_100}, except that the hyperparameter priors are now set to $\mu_{\rm pl} \sim {\cal U}(-150,150)$ and $\log \sigma_{\rm pl} \sim {\cal N}(0,15)$, substantially wider than those in Table~\ref{Table_subsecII:priorsHB}.

\begin{figure}
\flushleft
 \includegraphics[width=\columnwidth]{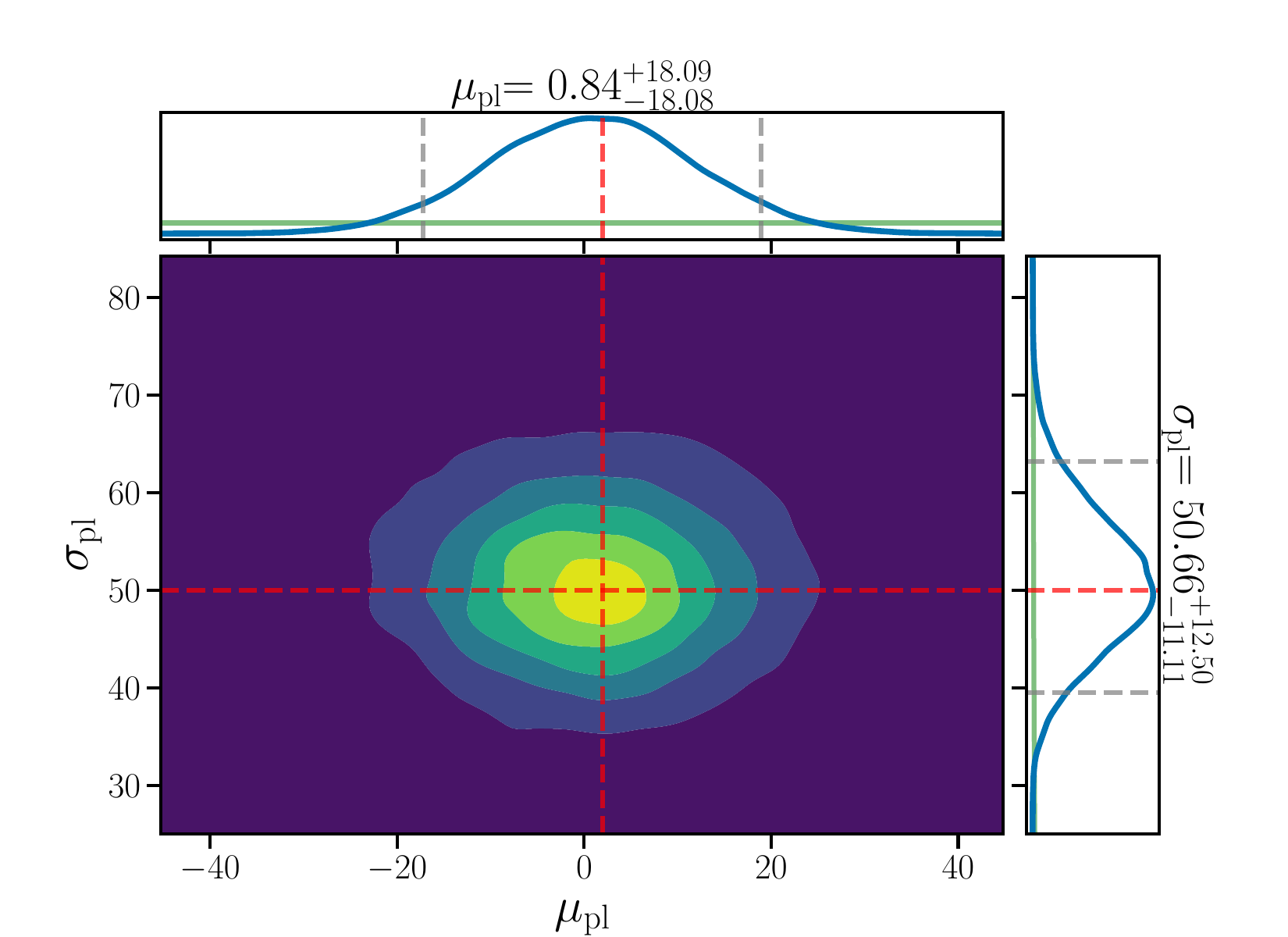}
 \label{fig_subsecII:corner_plot_Koft}
 \caption{Validation test for $\dot{K} \neq 0$ in the regime $| \dot{K}/ K | \gg | \dot{\nu} / \nu |$. Posterior distribution $p(\mu_{\rm pl}, \sigma_{\rm pl} \vert D)$ of the population-level braking index hyperparameters $\mu_{\rm pl}$ and $\sigma_{\rm pl}$, when drawing $n_{\rm pl}+\dot{K}_{\rm dim}$ from a wider distribution ${\cal N}(\mu_{\rm pl},\sigma_{\rm pl})$ than in Sections~\ref{subsec_III:results_M_100} and~\ref{subsec_III:varying_pop_size}, where $\dot{K}=0$ is assumed. Figures~\ref{fig_subsecII:corner_plot_100_pulsars} and \ref{fig_subsecII:corner_plot_Koft} share the same format. The vertical and horizontal red lines indicate the injected values $\mu_{\rm pl}=2$ and $\sigma_{\rm pl}=50$, respectively. The inferred medians of $\mu_{\rm pl}$ and $\sigma_{\rm pl}$ are displaced from the injected values by 58 and 1.3 per cent, respectively [equation (\ref{eq_subsecIII:error_theta})], corresponding to absolute displacements of $-1.2$ and $+0.66$, respectively. The injected values of $\mu_{\rm pl}$ and $\sigma_{\rm pl}$ fall within the inferred $90\%$ credible intervals.}

\end{figure}

Figure~\ref{fig_subsecII:corner_plot_Koft} presents the posterior distribution $p(\mu_{\rm pl}, \sigma_{\rm pl} \vert D)$ [equation~(\ref{Eq_subsecII:marg_post_mu_sigma})] as a standard corner plot for the $\dot{K} \neq 0$~test. The presentation of Figure~\ref{fig_subsecII:corner_plot_Koft} is identical to Figure~\ref{fig_subsecII:corner_plot_100_pulsars}. From equation~(\ref{eq_subsecIII:error_theta}), we obtain ${\rm ERR}(\mu_{\rm pl})=5.8\times10^{-1}$ and ${\rm ERR}(\sigma_{\rm pl})=1.3\times10^{-2}$. The injected values of $\mu_{\rm pl}=2$ and $\sigma_{\rm}=50$ fall within the inferred $90\%$ credible intervals. The per-pulsar posteriors for $(n_{\rm pl}+\dot{K}_{\rm dim})^{(m)}$ and $\chi^{(m)}$ contain $93\%$ and $78\%$, respectively, of the injected values within their $90\%$ credible intervals. Comparable accuracy is achieved when choosing an even wider prior for $\mu_{\rm pl}$, for example $\mu_{\rm pl} 
\sim {\cal U}(-10^{3},10^{3})$.

The validation of the hierarchical Bayesian scheme for $\dot{K} \neq 0$ and $| \dot{K} / K | \gg | \dot{\nu} / \nu |$ is an important result, because for real astronomical data there is no way to be sure if one has  $\dot{K}=0$ or $\dot{K} \neq 0$. Hence, when applying the method to real astronomical data, the hyperparameters prior $\pi(\mu_{\rm pl}, \sigma_{\rm pl})$ should be chosen wide enough to accommodate physically plausible values of $\dot{K}_{\rm dim} \neq 0$.

\section{Conclusions}
\label{Sec:Conclusions}

In this paper, we combine the Brownian model for anomalous braking indices described by equations~(\ref{Eq_secII:dX})--(\ref{Eq_SecII:dB_mem_less}) with a hierarchical Bayesian scheme to infer the posterior distribution $p(\mu_{\rm pl},\sigma_{\rm pl} | D)$ of the hyperparameters defining the population-level prior $\pi[n_{\rm pl}^{(m)},\chi^{(m)} | \mu_{\rm pl}, \sigma_{\rm pl} ] \propto {\cal N}(\mu_{\rm pl},\sigma_{\rm pl})$, as well as the per-pulsar posterior distribution $p[n_{\rm pl}^{(m)}, \chi^{(m)} | D]$, for a pulsar population of size $M$. The formal structure and practical implementation of the Bayesian scheme are described in Section~\ref{Sec: Methods (TBC)} and Appendix~\ref{AppendixA:HB_model}. 

The Bayesian scheme is validated through tests involving synthetic data to systematically quantify its accuracy under controlled conditions, i.e. when the injected parameters are known (see Sections~\ref{Sec:Results (TBC)}--\ref{subsec_III:varying_pop_size}). For test populations which approximately match the number of pulsars in previous studies of anomalous braking indices~\citep{JohnstonGalloway1999,ParthasarathyJohnston2020,LowerBailes2020}, viz. $M \gtrsim 50$, the hierarchical Bayesian scheme accurately recovers $\mu_{\rm pl}$ and $\sigma_{\rm pl}$; typically, their injected values are contained within their respective $90\%$ credible intervals. The accuracy of the scheme increases with $M$, ranging from $6.3\%$ ($M=50$) to $2\%$ ($M=300$) for $\mu_{\rm pl}$. On a per-pulsar basis, the percentage of pulsars whose $p[n^{(m)}_{\rm pl}\vert D]$ posteriors include the injected $n_{\rm pl}^{(m)}$ within their $90\%$ credible intervals grows from $66\%$ for $M=50$ to $87\%$ for $M \geq 100$. Posterior predictive checks confirm that the recovered and injected forms of the population-level braking index distribution $p[n_{\rm pl}^{\rm (pop)}]$ [equation (\ref{Eq_subsecII:marg_post_n_pop})] are in accord. We also confirm that the scheme is insensitive to the widths of the priors and performs well in the regime $| \dot{K} / K | \gg | \dot{\nu} / \nu |$, where one simply replaces $n_{\rm pl}$ by $n_{\rm pl} + \dot{K}_{\rm dim}$, an exact generalization \citep{VargasMelatos2024}. For a sample of $M=100$ synthetic pulsars, with braking indices $n_{\rm pl}+\dot{K}_{\rm dim}$ drawn from the relatively wide distribution ${\cal N}(\mu_{\rm pl}=2,\sigma_{\rm pl}=50)$, we find that the scheme accurately recovers $\mu_{\rm pl}$ and $\sigma_{\rm pl}$ to within $58\%$ and $1.3\%$ respectively, and the injected values are contained within their respective $90\%$ credible intervals. On a per-pulsar basis, the posteriors for $(n_{\rm pl}+\dot{K}_{\rm dim})^{(m)}$ and $\chi^{(m)}$ contain $93\%$ and $78\%$, respectively, of the injected values within their $90\%$ credible intervals, comparable to the corresponding fractions for $\dot{K}=0$.

The hierarchical Bayesian scheme in Section~\ref{Sec: Methods (TBC)} is straightforward to generalize, as the need arises. For example, the per-pulsar likelihood and population-level prior analyzed in this paper assume that all $M$ pulsars spin down secularly via the same mechanism, e.g. electromagnetic braking or gravitational radiation reaction (see Section~\ref{subsecII:chifromrms}). This assumption may not hold in reality. If some pulsars are dominated by electromagnetic braking and others by gravitational radiation reaction, then it is routine to replace equation~(\ref{Eq_secII:npl_distribution}) with a bimodal distribution.

Having validated the Bayesian scheme in Section~\ref{Sec: Methods (TBC)} with synthetic data, the next step is to apply it to real astronomical data in collaboration with the pulsar timing community. Several excellent data sets exist for this purpose, among them the sample of $85$ radio pulsars with high spin-down energy loss rate, $\dot{E}$, observed for $\gtrsim 10$ years with the $64$-m Parkes radio telescope~\citep{ParthasarathyShannon2019,ParthasarathyJohnston2020}. We postpone the analysis of real data to a forthcoming paper.

\section*{Acknowledgements}

The authors thank Patrick Meyers, for making the {\tt baboo} package freely available. Additionally, we thank Liam Dunn, Tom Kimpson, and Joe O'Leary for useful discussions regarding numerical methods which improved Section~\ref{subsecII:chifromrms} and Appendix~\ref{AppendixB}. We thank the anonymous referee for a constructive review, which prompted the addition of Section~\ref{sec:Koft}. This research was supported by the Australian Research Council Centre of Excellence for Gravitational Wave Discovery (OzGrav), grant number CE170100004. The numerical calculations were performed on the OzSTAR supercomputer facility at Swinburne University of Technology. The OzSTAR program receives funding in part from the Astronomy National Collaborative Research Infrastructure Strategy (NCRIS) allocation provided by the Australian Government.

\section*{Data availability}

All the synthetic data are generated using the open access software package {\tt baboo} available at \url{http://www.github.com/meyers-academic/baboo}~\citep{MeyersO'Neill2021}. We use~\tempoDOS~\citep{HobbsEdwards2006} to analyze the synthetic data.


\twocolumn

\bibliographystyle{mnras}
\bibliography{ADSABS_bib, main_non_ads_bib}

\newcommand{\noop}[1]{}
\begin{thebibliography}{}
\makeatletter
\relax
\def\mn@urlcharsother{\let\do\@makeother \do\$\do\&\do\#\do\^\do\_\do\%\do\~}
\def\mn@doi{\begingroup\mn@urlcharsother \@ifnextchar [ {\mn@doi@}
  {\mn@doi@[]}}
\def\mn@doi@[#1]#2{\def\@tempa{#1}\ifx\@tempa\@empty \href
  {http://dx.doi.org/#2} {doi:#2}\else \href {http://dx.doi.org/#2} {#1}\fi
  \endgroup}
\def\mn@eprint#1#2{\mn@eprint@#1:#2::\@nil}
\def\mn@eprint@arXiv#1{\href {http://arxiv.org/abs/#1} {{\tt arXiv:#1}}}
\def\mn@eprint@dblp#1{\href {http://dblp.uni-trier.de/rec/bibtex/#1.xml}
  {dblp:#1}}
\def\mn@eprint@#1:#2:#3:#4\@nil{\def\@tempa {#1}\def\@tempb {#2}\def\@tempc
  {#3}\ifx \@tempc \@empty \let \@tempc \@tempb \let \@tempb \@tempa \fi \ifx
  \@tempb \@empty \def\@tempb {arXiv}\fi \@ifundefined
  {mn@eprint@\@tempb}{\@tempb:\@tempc}{\expandafter \expandafter \csname
  mn@eprint@\@tempb\endcsname \expandafter{\@tempc}}}

\bibitem[\protect\citeauthoryear{{Abolmasov}, {Biryukov}  \&
  {Popov}}{{Abolmasov} et~al.}{2024}]{AbolmasovBiryukov2024}
{Abolmasov} P.,  {Biryukov} A.,   {Popov} S.~B.,  2024, \mn@doi [Galaxies]
  {10.3390/galaxies12010007}, \href
  {https://ui.adsabs.harvard.edu/abs/2024Galax..12....7A} {12, 7}

\bibitem[\protect\citeauthoryear{{Alpar} \& {Baykal}}{{Alpar} \&
  {Baykal}}{2006}]{AlparBaykal2006}
{Alpar} M.~A.,  {Baykal} A.,  2006, \mn@doi [\mnras]
  {10.1111/j.1365-2966.2006.10893.x}, \href
  {https://ui.adsabs.harvard.edu/abs/2006MNRAS.372..489A} {372, 489}

\bibitem[\protect\citeauthoryear{{Andersson}}{{Andersson}}{1998}]{Andersson1998}
{Andersson} N.,  1998, \mn@doi [\apj] {10.1086/305919}, \href
  {https://ui.adsabs.harvard.edu/abs/1998ApJ...502..708A} {502, 708}

\bibitem[\protect\citeauthoryear{{Antonelli}, {Basu}  \& {Haskell}}{{Antonelli}
  et~al.}{2022}]{AntonelliBasu2022}
{Antonelli} M.,  {Basu} A.,   {Haskell} B.,  2022, arXiv e-prints, \href
  {https://ui.adsabs.harvard.edu/abs/2022arXiv220610416A} {p. arXiv:2206.10416}

\bibitem[\protect\citeauthoryear{{Araujo}, {De Lorenci}, {Peter}  \&
  {Ruiz}}{{Araujo} et~al.}{2024}]{AraujoDeLorenci2024}
{Araujo} E.~C.~A.,  {De Lorenci} V.~A.,  {Peter} P.,   {Ruiz} L.~S.,  2024,
  \mn@doi [\mnras] {10.1093/mnras/stad3531}, \href
  {https://ui.adsabs.harvard.edu/abs/2024MNRAS.527.7956A} {527, 7956}

\bibitem[\protect\citeauthoryear{{Archibald} et~al.,}{{Archibald}
  et~al.}{2016}]{ArchibaldGotthelf2016}
{Archibald} R.~F.,  et~al., 2016, \mn@doi [\apjl]
  {10.3847/2041-8205/819/1/L16}, \href
  {https://ui.adsabs.harvard.edu/abs/2016ApJ...819L..16A} {819, L16}

\bibitem[\protect\citeauthoryear{{Arzoumanian}, {Nice}, {Taylor}  \&
  {Thorsett}}{{Arzoumanian} et~al.}{1994}]{ArzoumanianNice1994}
{Arzoumanian} Z.,  {Nice} D.~J.,  {Taylor} J.~H.,   {Thorsett} S.~E.,  1994,
  \mn@doi [\apj] {10.1086/173760}, \href
  {https://ui.adsabs.harvard.edu/abs/1994ApJ...422..671A} {422, 671}

\bibitem[\protect\citeauthoryear{{Barsukov} \& {Tsygan}}{{Barsukov} \&
  {Tsygan}}{2010}]{BarsukovTsygan2010}
{Barsukov} D.~P.,  {Tsygan} A.~I.,  2010, \mn@doi [\mnras]
  {10.1111/j.1365-2966.2010.17365.x}, \href
  {https://ui.adsabs.harvard.edu/abs/2010MNRAS.409.1077B} {409, 1077}

\bibitem[\protect\citeauthoryear{{Barsukov}, {Polyakova}  \&
  {Tsygan}}{{Barsukov} et~al.}{2009}]{BarsukovPolyakova2009}
{Barsukov} D.~P.,  {Polyakova} P.~I.,   {Tsygan} A.~I.,  2009, \mn@doi
  [Astronomy Reports] {10.1134/S1063772909120075}, \href
  {https://ui.adsabs.harvard.edu/abs/2009ARep...53.1146B} {53, 1146}

\bibitem[\protect\citeauthoryear{{Betancourt}}{{Betancourt}}{2017}]{Betancourt2017}
{Betancourt} M.,  2017, \mn@doi [arXiv e-prints] {10.48550/arXiv.1701.02434},
  \href {https://ui.adsabs.harvard.edu/abs/2017arXiv170102434B} {p.
  arXiv:1701.02434}

\bibitem[\protect\citeauthoryear{{Blandford} \& {Romani}}{{Blandford} \&
  {Romani}}{1988}]{BlandfordRomani1988}
{Blandford} R.~D.,  {Romani} R.~W.,  1988, \mn@doi [\mnras]
  {10.1093/mnras/234.1.57P}, \href
  {https://ui.adsabs.harvard.edu/abs/1988MNRAS.234P..57B} {234, 57P}

\bibitem[\protect\citeauthoryear{{Bransgrove}, {Levin}  \&
  {Beloborodov}}{{Bransgrove} et~al.}{2024}]{BransgroveLevin2024}
{Bransgrove} A.,  {Levin} Y.,   {Beloborodov} A.~M.,  2024, \mn@doi [arXiv
  e-prints] {10.48550/arXiv.2408.10888}, \href
  {https://ui.adsabs.harvard.edu/abs/2024arXiv240810888B} {p. arXiv:2408.10888}

\bibitem[\protect\citeauthoryear{{Bucciantini}, {Thompson}, {Arons}, {Quataert}
   \& {Del Zanna}}{{Bucciantini} et~al.}{2006}]{BucciantiniThompson2006}
{Bucciantini} N.,  {Thompson} T.~A.,  {Arons} J.,  {Quataert} E.,   {Del Zanna}
  L.,  2006, \mn@doi [\mnras] {10.1111/j.1365-2966.2006.10217.x}, \href
  {https://ui.adsabs.harvard.edu/abs/2006MNRAS.368.1717B} {368, 1717}

\bibitem[\protect\citeauthoryear{{Chukwude} \& {Chidi Odo}}{{Chukwude} \&
  {Chidi Odo}}{2016}]{ChukwudeChidiOdo2016}
{Chukwude} A.~E.,  {Chidi Odo} F.,  2016, \mn@doi [Research in Astronomy and
  Astrophysics] {10.1088/1674-4527/16/10/150}, \href
  {https://ui.adsabs.harvard.edu/abs/2016RAA....16..150C} {16, 150}

\bibitem[\protect\citeauthoryear{{Chukwude}, {Baiden}  \&
  {Onuchukwu}}{{Chukwude} et~al.}{2010}]{ChukwudeBaiden2010}
{Chukwude} A.~E.,  {Baiden} A.~A.,   {Onuchukwu} C.~C.,  2010, \mn@doi [\aap]
  {10.1051/0004-6361/200911634}, \href
  {https://ui.adsabs.harvard.edu/abs/2010A&A...515A..21C} {515, A21}

\bibitem[\protect\citeauthoryear{{Coles}, {Hobbs}, {Champion}, {Manchester}  \&
  {Verbiest}}{{Coles} et~al.}{2011}]{ColesHobbs2011}
{Coles} W.,  {Hobbs} G.,  {Champion} D.~J.,  {Manchester} R.~N.,   {Verbiest}
  J.~P.~W.,  2011, \mn@doi [\mnras] {10.1111/j.1365-2966.2011.19505.x}, \href
  {https://ui.adsabs.harvard.edu/abs/2011MNRAS.418..561C} {418, 561}

\bibitem[\protect\citeauthoryear{{Contopoulos} \& {Spitkovsky}}{{Contopoulos}
  \& {Spitkovsky}}{2006}]{ContopoulosSpitkovsky2006}
{Contopoulos} I.,  {Spitkovsky} A.,  2006, \mn@doi [\apj] {10.1086/501161},
  \href {https://ui.adsabs.harvard.edu/abs/2006ApJ...643.1139C} {643, 1139}

\bibitem[\protect\citeauthoryear{{Cook}, {Gelman}  \& Rubin}{{Cook}
  et~al.}{2006}]{Cook2006}
{Cook} S.~R.,  {Gelman} A.,   Rubin R. D.~B.,  2006, \mn@doi [Journal of
  Computational and Graphical Statistics] {10.1198/106186006X136976}, 15, 675

\bibitem[\protect\citeauthoryear{{Cordes}}{{Cordes}}{1980}]{Cordes1980}
{Cordes} J.~M.,  1980, \mn@doi [\apj] {10.1086/157861}, \href
  {https://ui.adsabs.harvard.edu/abs/1980ApJ...237..216C} {237, 216}

\bibitem[\protect\citeauthoryear{{Cordes} \& {Downs}}{{Cordes} \&
  {Downs}}{1985}]{CordesDowns1985}
{Cordes} J.~M.,  {Downs} G.~S.,  1985, \mn@doi [\apjs] {10.1086/191076}, \href
  {https://ui.adsabs.harvard.edu/abs/1985ApJS...59..343C} {59, 343}

\bibitem[\protect\citeauthoryear{{Cordes} \& {Helfand}}{{Cordes} \&
  {Helfand}}{1980}]{CordesHelfand1980}
{Cordes} J.~M.,  {Helfand} D.~J.,  1980, \mn@doi [\apj] {10.1086/158150}, \href
  {https://ui.adsabs.harvard.edu/abs/1980ApJ...239..640C} {239, 640}

\bibitem[\protect\citeauthoryear{{Faucher-Gigu{\`e}re} \&
  {Kaspi}}{{Faucher-Gigu{\`e}re} \& {Kaspi}}{2006}]{Faucher-GiguereKaspi2006}
{Faucher-Gigu{\`e}re} C.-A.,  {Kaspi} V.~M.,  2006, \mn@doi [\apj]
  {10.1086/501516}, \href
  {https://ui.adsabs.harvard.edu/abs/2006ApJ...643..332F} {643, 332}

\bibitem[\protect\citeauthoryear{{Gelman}, {Carlin}, {Stern}, {Dunson},
  {Vehtari}  \& {Rubin}}{{Gelman} et~al.}{2013}]{gelman2013bayesian}
{Gelman} A.,  {Carlin} J.~B.,  {Stern} H.~S.,  {Dunson} D.~B.,  {Vehtari} A.,
  {Rubin} D.~B.,  2013, Bayesian Data Analysis, 3rd edn.
Chapman and Hall/CRC, \mn@doi{10.1201/b16018}, \url
  {https://doi.org/10.1201/b16018}

\bibitem[\protect\citeauthoryear{{Goldreich}}{{Goldreich}}{1970}]{Goldreich1970}
{Goldreich} P.,  1970, \mn@doi [\apjl] {10.1086/180513}, \href
  {https://ui.adsabs.harvard.edu/abs/1970ApJ...160L..11G} {160, L11}

\bibitem[\protect\citeauthoryear{{Goncharov} et~al.,}{{Goncharov}
  et~al.}{2021}]{GoncharovReardon2021}
{Goncharov} B.,  et~al., 2021, \mn@doi [\mnras] {10.1093/mnras/staa3411}, \href
  {https://ui.adsabs.harvard.edu/abs/2021MNRAS.502..478G} {502, 478}

\bibitem[\protect\citeauthoryear{{G{\"u}gercino{\u{g}}lu}}{{G{\"u}gercino{\u{g}}lu}}{2017}]{Gugercinoglu2017}
{G{\"u}gercino{\u{g}}lu} E.,  2017, \mn@doi [\mnras] {10.1093/mnras/stx985},
  \href {https://ui.adsabs.harvard.edu/abs/2017MNRAS.469.2313G} {469, 2313}

\bibitem[\protect\citeauthoryear{{G{\"u}gercino{\u{g}}lu} \&
  {Alpar}}{{G{\"u}gercino{\u{g}}lu} \& {Alpar}}{2014}]{GugercinogluAlpar2014}
{G{\"u}gercino{\u{g}}lu} E.,  {Alpar} M.~A.,  2014, \mn@doi [\apjl]
  {10.1088/2041-8205/788/1/L11}, \href
  {https://ui.adsabs.harvard.edu/abs/2014ApJ...788L..11G} {788, L11}

\bibitem[\protect\citeauthoryear{{G{\"u}gercino{\u{g}}lu} \&
  {Alpar}}{{G{\"u}gercino{\u{g}}lu} \& {Alpar}}{2020}]{GugercinogluAlpar2020}
{G{\"u}gercino{\u{g}}lu} E.,  {Alpar} M.~A.,  2020, \mn@doi [\mnras]
  {10.1093/mnras/staa1672}, \href
  {https://ui.adsabs.harvard.edu/abs/2020MNRAS.496.2506G} {496, 2506}

\bibitem[\protect\citeauthoryear{{Gunn} \& {Ostriker}}{{Gunn} \&
  {Ostriker}}{1969}]{GunnOstriker1969}
{Gunn} J.~E.,  {Ostriker} J.~P.,  1969, \mn@doi [\nat] {10.1038/221454a0},
  \href {https://ui.adsabs.harvard.edu/abs/1969Natur.221..454G} {221, 454}

\bibitem[\protect\citeauthoryear{{Hobbs}, {Edwards}  \& {Manchester}}{{Hobbs}
  et~al.}{2006}]{HobbsEdwards2006}
{Hobbs} G.~B.,  {Edwards} R.~T.,   {Manchester} R.~N.,  2006, \mn@doi [\mnras]
  {10.1111/j.1365-2966.2006.10302.x}, \href
  {https://ui.adsabs.harvard.edu/abs/2006MNRAS.369..655H} {369, 655}

\bibitem[\protect\citeauthoryear{{Hobbs}, {Lyne}  \& {Kramer}}{{Hobbs}
  et~al.}{2010}]{HobbsLyne2010}
{Hobbs} G.,  {Lyne} A.~G.,   {Kramer} M.,  2010, \mn@doi [\mnras]
  {10.1111/j.1365-2966.2009.15938.x}, \href
  {https://ui.adsabs.harvard.edu/abs/2010MNRAS.402.1027H} {402, 1027}

\bibitem[\protect\citeauthoryear{{Johnston} \& {Galloway}}{{Johnston} \&
  {Galloway}}{1999}]{JohnstonGalloway1999}
{Johnston} S.,  {Galloway} D.,  1999, \mn@doi [\mnras]
  {10.1046/j.1365-8711.1999.02737.x}, \href
  {https://ui.adsabs.harvard.edu/abs/1999MNRAS.306L..50J} {306, L50}

\bibitem[\protect\citeauthoryear{{Johnston} \& {Karastergiou}}{{Johnston} \&
  {Karastergiou}}{2017}]{JohnstonKarastergiou2017}
{Johnston} S.,  {Karastergiou} A.,  2017, \mn@doi [\mnras]
  {10.1093/mnras/stx377}, \href
  {https://ui.adsabs.harvard.edu/abs/2017MNRAS.467.3493J} {467, 3493}

\bibitem[\protect\citeauthoryear{{Jones}}{{Jones}}{1990}]{Jones1990}
{Jones} P.~B.,  1990, \mnras, \href
  {https://ui.adsabs.harvard.edu/abs/1990MNRAS.246..364J} {246, 364}

\bibitem[\protect\citeauthoryear{{Keith} \& {Ni{\c{t}}u}}{{Keith} \&
  {Ni{\c{t}}u}}{2023}]{KeithNitu2023}
{Keith} M.~J.,  {Ni{\c{t}}u} I.~C.,  2023, \mn@doi [\mnras]
  {10.1093/mnras/stad1713}, \href
  {https://ui.adsabs.harvard.edu/abs/2023MNRAS.523.4603K} {523, 4603}

\bibitem[\protect\citeauthoryear{{Kou} \& {Tong}}{{Kou} \&
  {Tong}}{2015}]{KouTong2015}
{Kou} F.~F.,  {Tong} H.,  2015, \mn@doi [\mnras] {10.1093/mnras/stv734}, \href
  {https://ui.adsabs.harvard.edu/abs/2015MNRAS.450.1990K} {450, 1990}

\bibitem[\protect\citeauthoryear{{Lentati}, {Alexander}, {Hobson}, {Taylor},
  {Gair}, {Balan}  \& {van Haasteren}}{{Lentati}
  et~al.}{2013}]{LentatiAlexander2013}
{Lentati} L.,  {Alexander} P.,  {Hobson} M.~P.,  {Taylor} S.,  {Gair} J.,
  {Balan} S.~T.,   {van Haasteren} R.,  2013, \mn@doi [\prd]
  {10.1103/PhysRevD.87.104021}, \href
  {https://ui.adsabs.harvard.edu/abs/2013PhRvD..87j4021L} {87, 104021}

\bibitem[\protect\citeauthoryear{{Lentati}, {Alexander}, {Hobson}, {Feroz},
  {van Haasteren}, {Lee}  \& {Shannon}}{{Lentati}
  et~al.}{2014}]{LentatiAlexander2014}
{Lentati} L.,  {Alexander} P.,  {Hobson} M.~P.,  {Feroz} F.,  {van Haasteren}
  R.,  {Lee} K.~J.,   {Shannon} R.~M.,  2014, \mn@doi [\mnras]
  {10.1093/mnras/stt2122}, \href
  {https://ui.adsabs.harvard.edu/abs/2014MNRAS.437.3004L} {437, 3004}

\bibitem[\protect\citeauthoryear{{Link} \& {Epstein}}{{Link} \&
  {Epstein}}{1997}]{LinkEpstein1997}
{Link} B.,  {Epstein} R.~I.,  1997, \mn@doi [\apjl] {10.1086/310549}, \href
  {https://ui.adsabs.harvard.edu/abs/1997ApJ...478L..91L} {478, L91}

\bibitem[\protect\citeauthoryear{{Livingstone} \& {Kaspi}}{{Livingstone} \&
  {Kaspi}}{2011}]{LivingstoneKaspi2011}
{Livingstone} M.~A.,  {Kaspi} V.~M.,  2011, \mn@doi [\apj]
  {10.1088/0004-637X/742/1/31}, \href
  {https://ui.adsabs.harvard.edu/abs/2011ApJ...742...31L} {742, 31}

\bibitem[\protect\citeauthoryear{{Livingstone}, {Kaspi}, {Gavriil},
  {Manchester}, {Gotthelf}  \& {Kuiper}}{{Livingstone}
  et~al.}{2007}]{LivingstoneKaspi2007}
{Livingstone} M.~A.,  {Kaspi} V.~M.,  {Gavriil} F.~P.,  {Manchester} R.~N.,
  {Gotthelf} E.~V.~G.,   {Kuiper} L.,  2007, \mn@doi [\apss]
  {10.1007/s10509-007-9320-3}, \href
  {https://ui.adsabs.harvard.edu/abs/2007Ap&SS.308..317L} {308, 317}

\bibitem[\protect\citeauthoryear{{Livingstone}, {Ng}, {Kaspi}, {Gavriil}  \&
  {Gotthelf}}{{Livingstone} et~al.}{2011}]{LivingstoneNg2011}
{Livingstone} M.~A.,  {Ng} C.~Y.,  {Kaspi} V.~M.,  {Gavriil} F.~P.,
  {Gotthelf} E.~V.,  2011, \mn@doi [\apj] {10.1088/0004-637X/730/2/66}, \href
  {https://ui.adsabs.harvard.edu/abs/2011ApJ...730...66L} {730, 66}

\bibitem[\protect\citeauthoryear{{Lower} et~al.,}{{Lower}
  et~al.}{2020}]{LowerBailes2020}
{Lower} M.~E.,  et~al., 2020, \mn@doi [\mnras] {10.1093/mnras/staa615}, \href
  {https://ui.adsabs.harvard.edu/abs/2020MNRAS.494..228L} {494, 228}

\bibitem[\protect\citeauthoryear{{Lower} et~al.,}{{Lower}
  et~al.}{2021}]{LowerJohnston2021}
{Lower} M.~E.,  et~al., 2021, \mn@doi [\mnras] {10.1093/mnras/stab2678}, \href
  {https://ui.adsabs.harvard.edu/abs/2021MNRAS.508.3251L} {508, 3251}

\bibitem[\protect\citeauthoryear{Lyne, Pritchard  \& Graham~Smith}{Lyne
  et~al.}{1993}]{LyneAG1993}
Lyne A.~G.,  Pritchard R.~S.,   Graham~Smith F.,  1993, \mn@doi [Monthly
  Notices of the Royal Astronomical Society] {10.1093/mnras/265.4.1003}, 265,
  1003

\bibitem[\protect\citeauthoryear{{Lyne}, {Hobbs}, {Kramer}, {Stairs}  \&
  {Stappers}}{{Lyne} et~al.}{2010}]{LyneHobbs2010}
{Lyne} A.,  {Hobbs} G.,  {Kramer} M.,  {Stairs} I.,   {Stappers} B.,  2010,
  \mn@doi [Science] {10.1126/science.1186683}, \href
  {https://ui.adsabs.harvard.edu/abs/2010Sci...329..408L} {329, 408}

\bibitem[\protect\citeauthoryear{{Lyne}, {Jordan}, {Graham-Smith}, {Espinoza},
  {Stappers}  \& {Weltevrede}}{{Lyne} et~al.}{2015}]{LyneJordan2015}
{Lyne} A.~G.,  {Jordan} C.~A.,  {Graham-Smith} F.,  {Espinoza} C.~M.,
  {Stappers} B.~W.,   {Weltevrede} P.,  2015, \mn@doi [\mnras]
  {10.1093/mnras/stu2118}, \href
  {https://ui.adsabs.harvard.edu/abs/2015MNRAS.446..857L} {446, 857}

\bibitem[\protect\citeauthoryear{{Manchester}, {Hobbs}, {Teoh}  \&
  {Hobbs}}{{Manchester} et~al.}{2005}]{ManchesterHobbs2005}
{Manchester} R.~N.,  {Hobbs} G.~B.,  {Teoh} A.,   {Hobbs} M.,  2005, \mn@doi
  [\aj] {10.1086/428488}, \href
  {https://ui.adsabs.harvard.edu/abs/2005AJ....129.1993M} {129, 1993}

\bibitem[\protect\citeauthoryear{{Melatos}}{{Melatos}}{1997}]{Melatos1997}
{Melatos} A.,  1997, \mn@doi [\mnras] {10.1093/mnras/288.4.1049}, \href
  {https://ui.adsabs.harvard.edu/abs/1997MNRAS.288.1049M} {288, 1049}

\bibitem[\protect\citeauthoryear{{Melatos}}{{Melatos}}{2000}]{Melatos2000}
{Melatos} A.,  2000, \mn@doi [\mnras] {10.1046/j.1365-8711.2000.03031.x}, \href
  {https://ui.adsabs.harvard.edu/abs/2000MNRAS.313..217M} {313, 217}

\bibitem[\protect\citeauthoryear{{Melatos} \& {Link}}{{Melatos} \&
  {Link}}{2014}]{MelatosLink2014}
{Melatos} A.,  {Link} B.,  2014, \mn@doi [\mnras] {10.1093/mnras/stt1828},
  \href {https://ui.adsabs.harvard.edu/abs/2014MNRAS.437...21M} {437, 21}

\bibitem[\protect\citeauthoryear{{Melatos} \& {Millhouse}}{{Melatos} \&
  {Millhouse}}{2023}]{MelatosMillhouse2023}
{Melatos} A.,  {Millhouse} M.,  2023, \mn@doi [\apj]
  {10.3847/1538-4357/acbb6e}, \href
  {https://ui.adsabs.harvard.edu/abs/2023ApJ...948..106M} {948, 106}

\bibitem[\protect\citeauthoryear{{Meyers}, {Melatos}  \& {O'Neill}}{{Meyers}
  et~al.}{2021a}]{MeyersMelatos2021}
{Meyers} P.~M.,  {Melatos} A.,   {O'Neill} N.~J.,  2021a, \mn@doi [\mnras]
  {10.1093/mnras/stab262}, \href
  {https://ui.adsabs.harvard.edu/abs/2021MNRAS.502.3113M} {502, 3113}

\bibitem[\protect\citeauthoryear{{Meyers}, {O'Neill}, {Melatos}  \&
  {Evans}}{{Meyers} et~al.}{2021b}]{MeyersO'Neill2021}
{Meyers} P.~M.,  {O'Neill} N.~J.,  {Melatos} A.,   {Evans} R.~J.,  2021b,
  \mn@doi [\mnras] {10.1093/mnras/stab1952}, \href
  {https://ui.adsabs.harvard.edu/abs/2021MNRAS.506.3349M} {506, 3349}

\bibitem[\protect\citeauthoryear{{O'Neill}, {Meyers}  \& {Melatos}}{{O'Neill}
  et~al.}{2024}]{O'NeillMeyers2024}
{O'Neill} N.~J.,  {Meyers} P.~M.,   {Melatos} A.,  2024, arXiv e-prints, \href
  {https://ui.adsabs.harvard.edu/abs/2024arXiv240316467O} {p. arXiv:2403.16467}

\bibitem[\protect\citeauthoryear{{Onuchukwu} \& {Legahara}}{{Onuchukwu} \&
  {Legahara}}{2024}]{OnuchukwuLegahara2024}
{Onuchukwu} C.~C.,  {Legahara} E.,  2024, \mn@doi [\apss]
  {10.1007/s10509-024-04317-3}, \href
  {https://ui.adsabs.harvard.edu/abs/2024Ap&SS.369...51O} {369, 51}

\bibitem[\protect\citeauthoryear{{Owen}, {Lindblom}, {Cutler}, {Schutz},
  {Vecchio}  \& {Andersson}}{{Owen} et~al.}{1998}]{OwenLindblom1998}
{Owen} B.~J.,  {Lindblom} L.,  {Cutler} C.,  {Schutz} B.~F.,  {Vecchio} A.,
  {Andersson} N.,  1998, \mn@doi [\prd] {10.1103/PhysRevD.58.084020}, \href
  {https://ui.adsabs.harvard.edu/abs/1998PhRvD..58h4020O} {58, 084020}

\bibitem[\protect\citeauthoryear{{Papaloizou} \& {Pringle}}{{Papaloizou} \&
  {Pringle}}{1978}]{PapaloizouPringle1978}
{Papaloizou} J.,  {Pringle} J.~E.,  1978, \mn@doi [\mnras]
  {10.1093/mnras/182.3.423}, \href
  {https://ui.adsabs.harvard.edu/abs/1978MNRAS.182..423P} {182, 423}

\bibitem[\protect\citeauthoryear{{Parthasarathy} et~al.,}{{Parthasarathy}
  et~al.}{2019}]{ParthasarathyShannon2019}
{Parthasarathy} A.,  et~al., 2019, \mn@doi [\mnras] {10.1093/mnras/stz2383},
  \href {https://ui.adsabs.harvard.edu/abs/2019MNRAS.489.3810P} {489, 3810}

\bibitem[\protect\citeauthoryear{{Parthasarathy} et~al.,}{{Parthasarathy}
  et~al.}{2020}]{ParthasarathyJohnston2020}
{Parthasarathy} A.,  et~al., 2020, \mn@doi [\mnras] {10.1093/mnras/staa882},
  \href {https://ui.adsabs.harvard.edu/abs/2020MNRAS.494.2012P} {494, 2012}

\bibitem[\protect\citeauthoryear{{P{\'e}tri}}{{P{\'e}tri}}{2015}]{Petri2015}
{P{\'e}tri} J.,  2015, \mn@doi [\mnras] {10.1093/mnras/stv598}, \href
  {https://ui.adsabs.harvard.edu/abs/2015MNRAS.450..714P} {450, 714}

\bibitem[\protect\citeauthoryear{{P{\'e}tri}}{{P{\'e}tri}}{2017}]{Petri2017}
{P{\'e}tri} J.,  2017, \mn@doi [\mnras] {10.1093/mnras/stx2147}, \href
  {https://ui.adsabs.harvard.edu/abs/2017MNRAS.472.3304P} {472, 3304}

\bibitem[\protect\citeauthoryear{{Pons}, {Vigan{\`o}}  \& {Geppert}}{{Pons}
  et~al.}{2012}]{PonsVigano2012}
{Pons} J.~A.,  {Vigan{\`o}} D.,   {Geppert} U.,  2012, \mn@doi [\aap]
  {10.1051/0004-6361/201220091}, \href
  {https://ui.adsabs.harvard.edu/abs/2012A&A...547A...9P} {547, A9}

\bibitem[\protect\citeauthoryear{{Price}, {Link}, {Shore}  \& {Nice}}{{Price}
  et~al.}{2012}]{PriceLink2012}
{Price} S.,  {Link} B.,  {Shore} S.~N.,   {Nice} D.~J.,  2012, \mn@doi [\mnras]
  {10.1111/j.1365-2966.2012.21863.x}, \href
  {https://ui.adsabs.harvard.edu/abs/2012MNRAS.426.2507P} {426, 2507}

\bibitem[\protect\citeauthoryear{{Sedrakian} \& {Cordes}}{{Sedrakian} \&
  {Cordes}}{1998}]{SedrakianCordes1998}
{Sedrakian} A.,  {Cordes} J.~M.,  1998, \mn@doi [\apj] {10.1086/305896}, \href
  {https://ui.adsabs.harvard.edu/abs/1998ApJ...502..378S} {502, 378}

\bibitem[\protect\citeauthoryear{{Shannon} \& {Cordes}}{{Shannon} \&
  {Cordes}}{2010}]{ShannonCordes2010}
{Shannon} R.~M.,  {Cordes} J.~M.,  2010, \mn@doi [\apj]
  {10.1088/0004-637X/725/2/1607}, \href
  {https://ui.adsabs.harvard.edu/abs/2010ApJ...725.1607S} {725, 1607}

\bibitem[\protect\citeauthoryear{{Stairs} et~al.,}{{Stairs}
  et~al.}{2019}]{StairsLyne2019}
{Stairs} I.~H.,  et~al., 2019, \mn@doi [\mnras] {10.1093/mnras/stz647}, \href
  {https://ui.adsabs.harvard.edu/abs/2019MNRAS.485.3230S} {485, 3230}

\bibitem[\protect\citeauthoryear{{Tauris} \& {Konar}}{{Tauris} \&
  {Konar}}{2001}]{TaurisKonar2001}
{Tauris} T.~M.,  {Konar} S.,  2001, \mn@doi [\aap]
  {10.1051/0004-6361:20010988}, \href
  {https://ui.adsabs.harvard.edu/abs/2001A&A...376..543T} {376, 543}

\bibitem[\protect\citeauthoryear{{Thorne}}{{Thorne}}{1980}]{Thorne1980}
{Thorne} K.~S.,  1980, \mn@doi [Reviews of Modern Physics]
  {10.1103/RevModPhys.52.299}, \href
  {https://ui.adsabs.harvard.edu/abs/1980RvMP...52..299T} {52, 299}

\bibitem[\protect\citeauthoryear{{Urama}, {Link}  \& {Weisberg}}{{Urama}
  et~al.}{2006}]{UramaLink2006}
{Urama} J.~O.,  {Link} B.,   {Weisberg} J.~M.,  2006, \mn@doi [\mnras]
  {10.1111/j.1745-3933.2006.00192.x}, \href
  {https://ui.adsabs.harvard.edu/abs/2006MNRAS.370L..76U} {370, L76}

\bibitem[\protect\citeauthoryear{{Vargas} \& {Melatos}}{{Vargas} \&
  {Melatos}}{2023}]{VargasMelatos2023}
{Vargas} A.~F.,  {Melatos} A.,  2023, \mn@doi [\mnras]
  {10.1093/mnras/stad1301}, \href
  {https://ui.adsabs.harvard.edu/abs/2023MNRAS.522.4880V} {522, 4880}

\bibitem[\protect\citeauthoryear{{Vargas} \& {Melatos}}{{Vargas} \&
  {Melatos}}{2024}]{VargasMelatos2024}
{Vargas} A.~F.,  {Melatos} A.,  2024, \mn@doi [\mnras]
  {10.1093/mnras/stae2326}, \href
  {https://ui.adsabs.harvard.edu/abs/2024MNRAS.tmp.2267V} {534, 3410}

\bibitem[\protect\citeauthoryear{{Wasserman}, {Cordes}, {Chatterjee}  \&
  {Batra}}{{Wasserman} et~al.}{2022}]{WassermanCordes2022}
{Wasserman} I.,  {Cordes} J.~M.,  {Chatterjee} S.,   {Batra} G.,  2022, \mn@doi
  [\apj] {10.3847/1538-4357/ac38a6}, \href
  {https://ui.adsabs.harvard.edu/abs/2022ApJ...928...53W} {928, 53}

\bibitem[\protect\citeauthoryear{{Weltevrede}, {Johnston}  \&
  {Espinoza}}{{Weltevrede} et~al.}{2011}]{WeltevredeJohnston2011}
{Weltevrede} P.,  {Johnston} S.,   {Espinoza} C.~M.,  2011, \mn@doi [\mnras]
  {10.1111/j.1365-2966.2010.17821.x}, \href
  {https://ui.adsabs.harvard.edu/abs/2011MNRAS.411.1917W} {411, 1917}

\makeatother
\end{thebibliography}


\appendix

\section{Defining the statistical model} \label{AppendixA:HB_model}

A hierarchical model is appropriate when the system of interest can be subdivided into subsystems whose parameters are connected probabilistically, e.g. by being drawn from a single, universal, population-level distribution. In this appendix, we write down the general mathematical form of a hierarchical model and relate it to the specific system of interest in this paper --- $M$ braking index measurements for $M$ pulsars --- in order to justify the formulas in Section~\ref{subsecII:Bayesframework} and define the variables therein. The treatment is abridged; the reader is referred to Chapter 5 of~\cite{gelman2013bayesian} for a comprehensive discussion.

Let the system of interest (here, pulsars in the Milky Way) be subdivided into $M$ subsystems (here, $M$ individual pulsars), labelled by $1 \leq m \leq M$. Let the data associated with the $m$-th subsystem be $D^{(m)}$ and write $D=\{D^{(1)},\dots,D^{(M)}\}$. In this paper, we have $D^{(m)} = \{ n_{\rm meas}^{(m)}, \Delta n_{\rm meas}^{(m)}, S_{\rm meas}^{(m)}, \Delta S_{\rm meas}^{(m)}\}$; $n^{(m)}_{\rm meas}$ is the braking index measured for the $m$-th pulsar using equation~(\ref{Eq_Apndx2:n_diffstates}), $\Delta n_{\rm meas}^{(m)}$ is the associated measurement uncertainty, $S_{\rm meas}^{(m)}$ is the root mean square of the timing residuals defined by equation~(\ref{eq_subsecIII:S_meas_m}), and $\Delta S_{\rm meas}^{(m)}$ is the associated measurement uncertainty. Let the parameters associated with the $m$-th subsystem be $\theta^{(m)}$ and write $\theta=\{\theta^{(1)},\dots,\theta^{(M)}\}$. In this paper, we have $\theta^{(m)}=\{n^{(m)}_{\rm pl}, \chi^{(m)}\}$, i.e. the two unknowns on the right-hand side of equation~(\ref{Eq:Intro_variance_n_MES}). Assuming that the system is hierarchical, $\theta^{(m)}$ is drawn from a single, universal, population-level prior distribution $\pi[\theta^{(m)}\vert \psi]$, which is a function of a set of hyperparameters $\psi$. In this paper, we have $\psi=\{\mu_{\rm pl},\sigma_{\rm pl}\}$, and $\pi[\theta^{(m)}\vert \psi]$ is given by the Gaussian in equation~(\ref{Eq_secII:npl_distribution}). The hierarchical assumption contends, among other things, that all pulsars in the Milky Way share the same secular spin-down mechanism, described by a unimodal (Gaussian) population-level prior distribution for $n^{(m)}_{\rm pl}$. This assumption is a reasonable starting point but it may not be true, if the electromagnetic torque dominates in some pulsars, and the gravitational radiation reaction torque dominates in others, for example.

Bayes's theorem for the joint posterior of the subsystem parameters $\theta$ and hyperparameters $\psi$ takes the form 

\begin{equation}
    p(\psi,\theta \vert D) = {\cal Z}^{-1}{\cal L}( D \vert \psi, \theta)\pi(\psi,\theta),
    \label{Eq_appA:Bayestheorem_hyp}
\end{equation}

with

\begin{equation}
    {\cal Z}=\int d\psi d\theta {\cal L}(D \vert \psi,\theta)\pi(\psi,\theta),
    \label{Eq_appA:Bayes_evidence_hyp}
\end{equation}

where ${\cal L}(D\vert \psi,\theta)$ and $\pi(\psi,\theta)$ are the joint likelihood and joint prior respectively.

Suppose now that the subsystem parameters are exchangeable in the sense defined in Chapter 5 of~\cite{gelman2013bayesian}. That is, there exists no information beyond the data themselves that allows one to distinguish $\theta^{(m)}$ from $\theta^{(m'\neq m)}$. In this paper, exchangeability is a reasonable starting point, if one assumes that every pulsar obeys the same braking index physics, as proposed above. Moreover, there is no experimental reason to think that $n^{(m)}_{\rm meas}$ is measured one way for some pulsars and another way for the rest. Assuming exchangeability, the joint likelihood and joint prior factorize as

\begin{align}
    {\cal L}(D\vert \psi,\theta) &= \prod_{m'=1}^{M} {\cal L}^{(m')}[D^{(m')}\vert \psi,\theta^{(m')}] \label{Eq_appA:stepA3}\\
    &=  \prod_{m'=1}^{M} {\cal L}^{(m')}[D^{(m')}\vert \theta^{(m')}] \label{eq_appA:stepA4} 
\end{align}

and 

\begin{equation}
    \pi(\psi,\theta) = \prod_{m''=1}^{M} \pi[\theta^{(m'')}\vert \psi]\pi(\psi),
    \label{eq_appA:prior_decomposition_exch}
\end{equation}

where ${\cal L}^{(m')}$ is the likelihood of the $m'$-th subsystem and $\pi(\psi)$ is the hyperparameter prior. Equation~(\ref{eq_appA:stepA4}) follows from equation~(\ref{Eq_appA:stepA3}), because the problem is hierarchical; ${\cal L}^{(m')}$ does not depend directly on $\psi$. The Brownian model in this paper (see Section~\ref{subsecII:Brownianmodel}) satisfies this property; the right-hand side of equation~(\ref{Eq:Intro_variance_n_MES}), which defines ${\cal L}^{(m')}$, depends on $n_{\rm pl}^{(m)}$ and $\chi^{(m)}$ but not $\mu_{\rm pl}$ and $\sigma_{\rm pl}$. With the factorization in equations (\ref{Eq_appA:stepA3})--(\ref{eq_appA:prior_decomposition_exch}), equation~(\ref{Eq_appA:Bayes_evidence_hyp}) reduces to 

\begin{equation}
    {\cal Z} = \int d\psi \, \pi(\psi) \prod_{m'=1}^{M} \int d\theta^{(m')} {\cal L}^{(m')}[D^{(m')}\vert \theta^{(m')}] \pi[\theta^{(m')}\vert \psi].
    \label{eq_appA:bayes_evidence_fact_exch}
\end{equation}

We are now in a position to calculate various marginalized posteriors of physical interest. Upon marginalizing equation~(\ref{Eq_appA:Bayestheorem_hyp}) over $\theta$, we obtain the posterior of the hyperparameters,

\begin{align}
    p(\psi \vert D) =& \int d \theta p(\psi,\theta \vert D) \label{eq_appA:A7} \\ 
    =&\,{\cal Z}^{-1} \int d\theta^{(1)}\dots d\theta^{(M)} \nonumber  \\
    &\times\prod_{m'=1}^{M} {\cal L}^{(m')}[D^{(m')}\vert \theta^{(m')}]\pi[\theta^{(m')}\vert \psi] \pi(\psi) \\
    =&\,{\cal Z}^{-1}  \pi(\psi) \nonumber  \\
    &\times\prod_{m'=1}^{M} \int d\theta^{(m')} {\cal L}^{(m')}[D^{(m')}\vert \theta^{(m')}] \pi[\theta^{(m')}\vert \psi]. \label{eq_appA:step_A9}
\end{align}

Likewise, upon marginalizing equation~(\ref{Eq_appA:Bayestheorem_hyp}) over $\psi$, we obtain the posterior of all the subsystem parameters,

\begin{align}
    p(\theta \vert D) &= \int d \psi p(\psi,\theta \vert D) \label{eq_appA:step_A10}\\
    &= {\cal Z}^{-1} \prod_{m'=1}^{M}  {\cal L}^{(m')}[D^{(m')}\vert \theta^{(m')}] \int d\psi \, \pi[\theta^{(m')}\vert \psi] \pi(\psi). \label{eq_appA:step_A11}
\end{align}

If there is physical interest specifically in the parameters $\theta^{(m)}$ of the $m$-th subsystem, e.g. $n^{(m)}_{\rm pl}$ and $\chi^{(m)}$ for the $m$-th pulsar, then the associated posterior can be calculated by marginalizing equations~(\ref{eq_appA:step_A10}) or (\ref{eq_appA:step_A11}) over all the other subsystem parameters, viz.

\begin{align}
    p[\theta^{(m)}\vert D] =& \int d\psi d\theta^{(1)} \dots d\theta^{(m-1)} \nonumber \\
    &\times \int d\theta^{(m+1)}\dots d\theta^{(M)} p(\psi,\theta \vert D).
    \label{eq_appA:posterior_mth_pulsar}
\end{align}

Equation~(\ref{eq_appA:posterior_mth_pulsar}) does not imply that $p[\theta^{(m)}\vert D]$ is proportional to ${\cal L}^{(m)}[D^{(m)}\vert \theta^{(m)}]\int d\psi\pi[\theta^{(m)}\vert \psi] \pi(\psi)$. That is, one cannot write down a ``naive'' form of Bayes's theorem at the subsystem level. Moreover, one cannot cancel factors in the products over $m'$ in equations~(\ref{eq_appA:step_A9}) and (\ref{eq_appA:step_A11}) with similar-looking factors in ${\cal Z}$ in equation~(\ref{eq_appA:bayes_evidence_fact_exch}), because the latter factors are nested inside the $\psi$ integral in equation~(\ref{eq_appA:bayes_evidence_fact_exch}).

Finally, we are interested physically in the population-level posterior distribution of the subsystem parameters without reference to the $m$-th specific subsystem. We write this posterior as $p[\theta^{({\rm pop})}\vert D]$, replacing $\theta^{(m)}$ with $\theta^{({\rm pop})}$ to emphasize that we are not specializing to a particular subsystem, and that $\theta^{(1)},\dots,\theta^{(M)}$ are exchangeable in the sense defined in Chapter 5 of~\cite{gelman2013bayesian}. In this paper, for example, we are interested in the posterior distribution of $n_{\rm pl}^{({\rm pop})}$, the power-law exponent in the secular braking law $\dot{\nu} \propto \nu^{n_{\rm pl}^{({\rm pop})}}$ across the whole population of Milky Way pulsars, as represented by the sample analyzed in this paper. The posterior $p[\theta^{({\rm pop})}\vert D]$ is calculated from the population-level prior and hyperparameter posterior according to

\begin{equation}
    p[\theta^{({\rm pop})}\vert D] = \int d\psi \, \pi[\theta^{(m)} \mapsto \theta^{({\rm pop})} \vert \psi] p(\psi \vert D),
    \label{eq_appA:step13}
\end{equation}

where $p(\psi \vert D)$ is given by equations (\ref{eq_appA:A7})--(\ref{eq_appA:step_A9}). Note that $p[\theta^{({\rm pop})}\vert D]$ and $p[\theta^{(m)}\vert D]$ are different quantities;  the latter refers specifically to the $m$-th pulsar, whereas the former does not. Likewise, the right-hand sides of equation~(\ref{eq_appA:posterior_mth_pulsar}) and (\ref{eq_appA:step13}) are not equal.

\section{Justification of the $S_{\rm \MakeLowercase{meas}}^{(\MakeLowercase{m})}$ dependence in the per-pulsar  likelihood}
\label{AppendixB}

A central ingredient in the Bayesian scheme described in Section~\ref{Sec: Methods (TBC)} is the factor featuring $S_{\rm meas}^{(m)}$ in the per-pulsar likelihood, i.e. equation~(\ref{Eq_secII:zeta}). In this appendix we show, through analytic calculations and simulations involving synthetic data, that $\log~S_{\rm meas}^{(m)}$ follows approximately a normal distribution. We also show how to calculate the parameters $\mu_{S \rm, BM}[\chi^{(m)}]$ and $\sigma_{S \rm,BM}$ in equation~(\ref{Eq_secII:zeta}).

We start with a set of synthetic TOAs $\{t_{1}^{(m)},...,t_{N^{(m)}_{\rm TOA}}^{(m)}\}$ generated from the Brownian model [equations~(\ref{Eq_secII:dX})--(\ref{Eq_SecII:dB_mem_less})] for input parameters which emulate the $m$-th pulsar, viz. $\{{\rm RA}^{(m)},{\rm DEC}^{(m)}, {\bf X}^{(m)}(t_{0}), T_{\rm obs}^{(m)}, N_{\rm TOA}^{(m)}, [\sigma^{(m)}_{\ddot{\nu}}]^{2}, \gamma_{\nu}, \gamma_{\dot{\nu}}, \gamma_{\ddot{\nu}}^{(m)}\}$. The TOAs are fitted using a polynomial ephemeris [up to $\ddot{\nu}(t_{\rm ref})$] over the observation time $T_{\rm obs}^{(m)}$. In practice, we use \tempoDOS~to fit the TOAs, but other timing software can be used just as well. The polynomial fit yields measurements of $\nu^{(m)}(t_{\rm ref})$ and its higher-order time derivatives, the timing residuals $\{ {\cal R}[t_1^{(m)}], \dots, {\cal R}[ t_{N_{\rm TOA}^{(m)}}^{(m)} ] \}$, and the associated uncertainties $\{ \Delta{\cal R}[t_1^{(m)}], \dots, \Delta{\cal R}[ t_{N_{\rm TOA}^{(m)}}^{(m)} ] \}$. Following previous authors~\citep{CordesHelfand1980,HobbsLyne2010,ShannonCordes2010,AntonelliBasu2022}, we adopt the standard deviation of the measured timing residuals, $[S^{(m)}_{\rm meas}]^{2}$, as an indicator of the timing noise strength, viz. equation~(\ref{eq_subsecIII:S_meas_m}).

Equation~(\ref{eq_subsecIII:S_meas_m}) represents a sample standard deviation. That is, it equals the standard deviation of the measured timing residuals for a single realization of the Brownian model. Another realization with the same $[\sigma^{(m)}_{\ddot{\nu}}]^{2}$ and a new random seed generates a unique, yet statistically equivalent, TOA sequence and hence $[S^{(m)}_{\rm meas}]^{2}$. To account for this, we approximate the Brownian model as a continuous-time process and calculate the ensemble's variance

\begin{equation}
    \langle [S^{(m)}_{\rm meas}]^{2} \rangle \approx \frac{1}{T_{\rm obs}^{(m)}} \int_{0}^{T_{\rm obs}^{(m)}} dt \langle [{\cal R}^{(m)}(t)]^{2} \rangle. \label{eq_subsecIII:ensemble_var_S_meas} 
\end{equation}

For the Brownian model, $\langle [{\cal R}^{(m)}(t)] \rangle$ is given by $\langle [{\cal R}^{(m)}(t)]^{2} \rangle = \langle [\delta \phi^{(m)}(t)]^{2} \rangle/[2\pi\nu^{(m)}(t_{\rm ref})]^{2}$, where $\delta \phi^{(m)}(t) = \phi^{(m)}(t) - \phi_{\rm pl}^{(m)}(t)$ executes zero-mean fluctuations, and we have $\phi_{\rm pl}^{(m)}(t) \approx \phi_{0}^{(m)}+\nu^{(m)}(t_{\rm ref})(t-t_{\rm ref})+\dot{\nu}^{(m)}(t_{\rm ref})(t-t_{\rm ref})^{2}/2+\ddot{\nu}^{(m)}(t_{\rm ref})(t-t_{\rm ref})^{3}/6$ [see Equation (A10) in~\cite{VargasMelatos2023}]. Upon combining $\langle [{\cal R}^{(m)}(t)]^{2} \rangle = \langle [\delta \phi^{(m)}(t)]^{2} \rangle/[2\pi\nu^{(m)}(t_{\rm ref})]^{2}$ with (\ref{eq_subsecIII:ensemble_var_S_meas}), and Fourier transforming, we can write

\begin{equation}
\langle [S^{(m)}_{\rm meas}]^{2} \rangle \approx \frac{2}{[2\pi\nu^{(m)}(t_{\rm ref})]^{2}}\int_{\omega_{\rm c}^{(m)}}^{\infty} \frac{d\omega}{2\pi} \langle \vert \delta \hat{\phi}^{(m)}(\omega) \vert^{2} \rangle, \label{eq:FIntegral_Smeas}
\end{equation}

where the hat denotes a Fourier transform, $\langle \vert \delta \hat{\phi}^{(m)}(\omega) \vert^{2} \rangle$ is the two-sided PSD for the phase residuals, given by

\begin{equation}
    \langle \vert \delta \hat{\phi}^{(m)}(\omega) \vert^{2} \rangle = \frac{[\sigma^{(m)}_{\ddot{\nu}}]^{2}}{\omega^{2}(\gamma_{\nu}^{2}+\omega^{2})(\gamma_{\dot{\nu}}^{2}+\omega^{2})([\gamma_{\ddot{\nu}}^{(m)}]^{2}+\omega^{2})}, \label{Eq:PSD_phase_BM}
\end{equation}

and $\omega_{c}^{(m)}=2\pi/T_{\rm obs}^{(m)}$ is an arbitrary cut-off used to avoid the low-frequency ($\omega \rightarrow 0$) divergence of $\langle \vert \delta \hat{\phi}^{(m)}(\omega) \vert^{2} \rangle$. The prefactor of two in (\ref{eq:FIntegral_Smeas}) accounts for the negative frequencies in $\langle \vert \delta \hat{\phi}^{(m)}(\omega) \vert^{2} \rangle$. Combining equations (\ref{eq:FIntegral_Smeas}) and (\ref{Eq:PSD_phase_BM}) and performing the right-hand integral in the astrophysically relevant regime $\gamma_\nu \sim \gamma_{\dot{\nu}} \ll [T_{\rm obs}^{(m)}]^{-1} \ll \gamma_{\ddot{\nu}}^{(m)}$ [see Appendix A1 of \cite{VargasMelatos2023}], we obtain  

\begin{equation}
    \langle [S^{(m)}_{\rm meas}]^{2} \rangle = \frac{[\chi^{(m)}]^{2}[T_{\rm obs}^{(m)}]^{5}}{160\pi^{6}[\nu^{(m)}(t_{\rm ref})]^{2}}, \label{Eq:final_ensemble_Smeas}
\end{equation}

with $[\chi^{(m)}]^{2} = [\sigma^{(m)}_{\ddot{\nu}}]^{2}/[\gamma_{\ddot{\nu}}^{(m)}]^{2}$. Note that $\nu^{(m)}(t_{\rm ref})$ and $T_{\rm obs}^{(m)}$ are measured accurately, as discussed in footnote 4.

To calibrate for the cut-off $\omega_{c}^{(m)}$, equation~(\ref{Eq:final_ensemble_Smeas}) should be multiplied in general by a dimensionless constant, $C_{0}$, to match~\tempoDOS's polynomial fits for an ensemble of synthetic realizations. To empirically determine $C_{0}$, we run the following calibration experiment. Firstly, we generate $10^{3}$ TOA realizations which emulate a synthetic pulsar (say the $m$-th one), with parameters $\{{\rm RA}^{(m)},{\rm DEC}^{(m)}, {\bf X}^{(m)}(t_{0}), T_{\rm obs}^{(m)}, N_{\rm TOA}^{(m)}, [\sigma^{(m)}_{\ddot{\nu}}]^{2}, \gamma_{\nu}, \gamma_{\dot{\nu}}, \gamma_{\ddot{\nu}}^{(m)}\}$. Secondly, we analyze each TOA sequence (indexed by $i$) with~\tempoDOS~to obtain $[S^{(m)}_{\rm meas}]^{2}_{i}$ for $1 \leq i \leq 10^{3}$ via (\ref{eq_subsecIII:S_meas_m}). The $10^{3}$ values of $[S^{(m)}_{\rm meas}]^{2}_{i}$ are averaged to obtain $\langle [S^{(m)}_{\rm meas}]^{2} \rangle_{\mathrm{{\scriptstyle T2}}}$. Thirdly, we repeat the first and second steps for different values of $[\chi^{(m)}]^{2}$ --- holding all the other parameters constant --- covering the range $10^{-42} \leq [\chi^{(m)}]^{2}/(1\;{\rm s}^{-5}) \leq 10^{-38}$ and obtain additional estimates of $\langle [S^{(m)}_{\rm meas}]^{2} \rangle_{\mathrm{{\scriptstyle T2}}}$. To finish, we compare the output of (\ref{Eq:final_ensemble_Smeas}) with the experimentally measured $\langle [S^{(m)}_{\rm meas}]^{2} \rangle_{\mathrm{{\scriptstyle T2}}}$ values to obtain $C_{0}$. We repeat this calibration experiment for different synthetic pulsars [in particular with different $T_{\rm obs}^{(m)}$ and $\nu^{(m)}(t_{\rm ref})$] and find $C_{0} \approx 6.3\times10^{-2}$ on average, i.e. $\langle [S^{(m)}_{\rm meas}]^{2} \rangle_{\mathrm{{\scriptstyle T2}}} \approx 6.3\times10^{-2} \langle [S^{(m)}_{\rm meas}]^{2} \rangle$. Similarly, we assume that the mean $\langle S^{(m)}_{\rm meas} \rangle$ is given by $\langle S^{(m)}_{\rm meas} \rangle = C_{1} \sqrt{\langle [S^{(m)}_{\rm meas}]^{2} \rangle}$. We repeat the same calibration experiment, replacing $\langle [S^{(m)}_{\rm meas}]^{2} \rangle_{\mathrm{{\scriptstyle T2}}}$ with $\langle S^{(m)}_{\rm meas}\rangle_{\mathrm{{\scriptstyle T2}}}$, and find $C_{1} \approx 4.5\times10^{-1}$ on average, i.e. $\langle S^{(m)}_{\rm meas}\rangle_{\mathrm{{\scriptstyle T2}}}=4.5\times10^{-1} \sqrt{\langle [S^{(m)}_{\rm meas}]^{2} \rangle}$. This calibration experiment should be repeated, if the timing software or the degree of the polynomial fit are changed.   

\begin{figure}
\flushleft
 \includegraphics[width=\columnwidth]{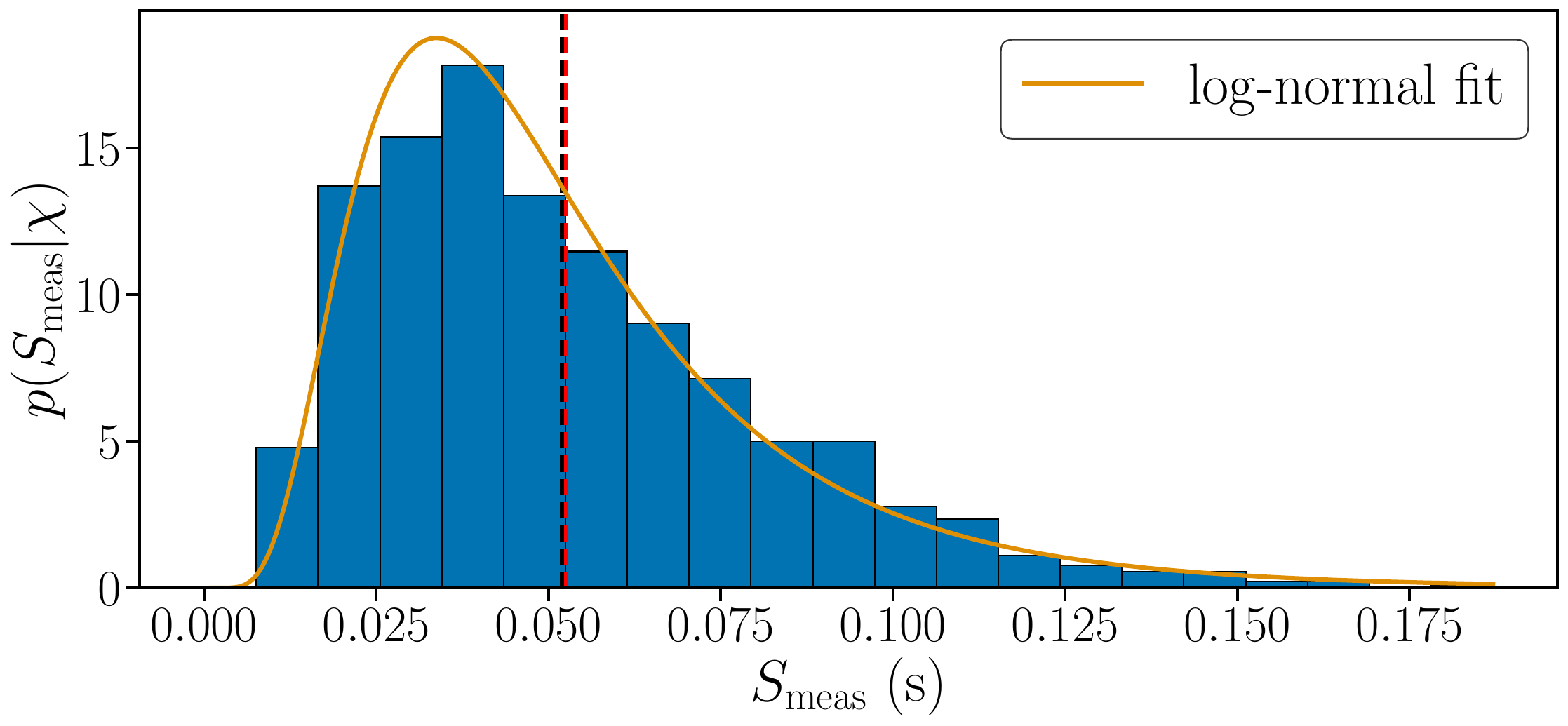}
 \caption{Distribution of $S_{\rm meas}$ measurements (blue histogram) for $10^{3}$ random realizations of synthetic data with the rotational parameters characteristic of PSR J0942$-$5552 from Table~\ref{Table_subsecII:example_injected_values} [superscript $(m)$ omitted temporarily]. The red-dashed line represents $\langle S_{\rm meas} \rangle_{\mathrm{{\scriptstyle T2}}}$, while the nearly overlapping black-dashed line represents $C_{1}\sqrt{\langle S_{\rm meas}^{2} \rangle}=5.2\times10^{-2}~{\rm s}$ from (\ref{Eq:final_ensemble_Smeas}) with $\chi_{\rm inj}^{2}=10^{-38}~{\rm s}^{-5}$, $C_{1}=4.5\times10^{-1}$, and the $\nu(t_{0})$ and $T_{\rm obs}$ parameters in Table~\ref{Table_subsecII:example_injected_values}. The fractional error between $\langle S_{\rm meas} \rangle_{\mathrm{{\scriptstyle T2}}}$ and $C_{1}\sqrt{\langle S_{\rm meas}^{2} \rangle}$ is less than $1\%$. The orange curve represents the log-normal fit obtained through (\ref{Eq_secII:zeta}) with parameters $\mu_{S \rm,BM}=-3.1$ [equation~(\ref{eq:mu_SBM})] and $\sigma_{S \rm,BM}^{2}=0.27$ [equation~(\ref{eq:sigma_SBM})]. The possible values of $S_{\rm meas}$ span $\approx 1.4$ dex even when all the realizations share the same injected values.}
\label{fig_subsecII:histogram_S_R_example}
\end{figure}

Once calibrated, equation~(\ref{Eq:final_ensemble_Smeas}) determines the mean and variance of the distribution of $S_{\rm meas}^{(m)}$ values for the $m$-th pulsar. To determine the distribution of $S_{\rm meas}^{(m)}$ values, for a fixed set of pulsar parameters, we fit a probability distribution to the $S_{\rm meas}^{(m)}$ histogram generated by the calibration experiment in the previous paragraph. Figure~\ref{fig_subsecII:histogram_S_R_example} shows an example, using the pulsar parameters recorded in Table~\ref{Table_subsecII:example_injected_values}, which emulate the representative object PSR J0942$-$5552 studied by~\cite{LowerBailes2020}~(we omit temporarily the index $m$). The histogram contains $10^{3}$ realizations of equations~(\ref{Eq_secII:dX})--(\ref{Eq_SecII:dB_mem_less}), which are converted into a measurement of $S_{\rm meas}$ [equation~(\ref{eq_subsecIII:S_meas_m})] after being analyzed using~\tempoDOS. The red, dashed, vertical line
 represents $\langle S_{\rm meas} \rangle_{\mathrm{{\scriptstyle T2}}}$, while the black, dashed, vertical line represents $C_{1}\sqrt{\langle S_{\rm meas}^{2} \rangle(\chi)}=5.2\times10^{-2}~{\rm s}$, obtained from equation~(\ref{Eq:final_ensemble_Smeas}) using the parameters in Table~\ref{Table_subsecII:example_injected_values}. The fractional error between $\langle S_{\rm meas} \rangle_{\mathrm{{\scriptstyle T2}}}$ and $C_{1}\sqrt{\langle S_{\rm meas}^{2}\rangle(\chi)}$ is less than $1\%$. The orange curve corresponds to the log-normal in equation~(\ref{Eq_secII:zeta}) with parameters 

 \begin{equation}
     \mu_{S \rm, BM}(\chi) = \log\Bigg[C_{1}\sqrt{\langle S_{\rm meas}^{2} \rangle(\chi)}\Bigg]-\frac{1}{2}\log\Bigg(1+\frac{C_{0}}{C_{1}^{2}}\Bigg), \label{eq:mu_SBM}
 \end{equation}

and 

\begin{equation}
    \sigma^{2}_{S \rm,BM} = \log\Bigg(1+\frac{C_{0}}{C_{1}^{2}}\Bigg).\label{eq:sigma_SBM}
\end{equation}

We note that equations~(\ref{eq:mu_SBM}) and~(\ref{eq:sigma_SBM}) are the mean and variance of $\log~S_{\rm meas}$. For the Brownian model, $\sigma^{2}_{S \rm,BM}$ is a constant, independent of the pulsar parameters, once the calibration experiments fix $C_{0}$ and $C_{1}$. Simulations using different pulsar parameters show that the standard deviation calculated via~\tempoDOS~is $\approx 0.5$, while equation~(\ref{eq:sigma_SBM}) yields $\sigma_{S \rm,BM}=0.52$ for $C_{0}=6.3\times10^{-2}$ and $C_{1}=4.5\times10^{-1}$. A Kolmogorov-Smirnov test between the $S_{\rm meas}$ data (blue histogram) and the log-normal fit yields a $p$-value of $0.12$. In general, other distributions can be fitted, such as a $\Gamma$-distribution. Simulations for various pulsar parameters show that the $p$-values returned by a Kolmogorov-Smirnov test depend weakly on the form of the distribution. Hence, for simplicity, we opt for the log-normal (\ref{Eq_secII:zeta}).

\begin{table}
\centering
\caption{Injected Brownian model parameters for the example of fitting the distribution of $S_{\rm meas}$ values generated at a fixed $\chi$~in Appendix~\ref{AppendixB} [superscript $(m)$ omitted temporarily]. The values of $\nu(t_{0})$ and $\dot{\nu}(t_{0})$ coincide with those of PSR J0942$-$5552, an arbitrary but representative pulsar studied by~\protect\cite{LowerBailes2020}. The injected value of $\ddot{\nu}(t_{0})$ implies $n_{\rm pl}=3$. The parameters in the lower half of the table are used to generate the synthetic data from the Brownian model, viz. (\ref{Eq_secII:dX})--(\ref{Eq_SecII:dB_mem_less}). The rotational and model parameters are consistent with an injected value $\chi_{\rm inj}^{2}=10^{-38}~{\rm s}^{-5}$.}
\label{Table_subsecII:example_injected_values}
\begin{tabularx}{\columnwidth}{XXX}
\hline
Parameter & Units & Injected value  \\
\hline
 $\nu(t_{0})$ & $\text{Hz}$ & $1.5051430406$ \\
 $\dot{\nu}(t_{0})$ & $10^{-14}~\text{Hz s}^{-1}$ & $-5.1380792001$ \\ 
 $\ddot{\nu}(t_{0})$ & $10^{-24}~\text{Hz s}^{-2}$ & $5.23\times10^{-3}$ \\ 
 \hline
 $\gamma_{\nu}$ & ${\rm s}^{-1}$ & $1\times10^{-13}$ \\ 
 $\gamma_{\dot{\nu}}$ & ${\rm s}^{-1}$ & $1\times10^{-13}$ \\
 $\gamma_{\ddot{\nu}}$ & ${\rm s}^{-1}$ & $1\times10^{-6}$ \\
$\sigma_{\ddot{\nu}}^{2}$ & ${\rm Hz}^{2}{\rm s}^{-5}$ & $1\times10^{-50}$ \\ 
$T_{\text{obs}}$ & $\text{days}$ & $2.5\times10^{3}$ \\
$N_{\text{TOA}}$ & -- & $1.5\times10^{2}$ \\ 
$\Delta_{\rm TOA}$ & $\mu \text{s}$ & $1\times10^{2}$ \\ 
\hline
\end{tabularx}
\end{table}

\bsp	
\label{lastpage}
\end{document}